\documentclass[11pt]{report}

\usepackage[dvipdfm]{graphicx}
\usepackage{amsmath}
\usepackage{bm}
\usepackage{fancyhdr}
\usepackage{makeidx}
\graphicspath{{fig/}}

\newcommand\beq{\begin{equation}}
\newcommand\eeq{\end{equation}}
\newcommand\beqa{\begin{eqnarray}}
\newcommand\eeqa{\end{eqnarray}}

\newcommand\bp{{\bf p}}
\newcommand\br{{\bf r}}
\newcommand\tsurf{{\tau_S}}

\makeatletter
\def\ps@fancy{%
\def\chaptermark##1{\markboth{\ifnum \c@secnumdepth>\z@
Chapter \thechapter\hskip 1em\relax \fi ##1}{}}%
\def\sectionmark##1{\markright{\ifnum \c@secnumdepth>\z@
\thesection\hskip 1em\relax \fi ##1}{}}%
\def\subsectionmark##1{\markright {\ifnum \c@secnumdepth >\@ne
\thesubsection\hskip 1em\relax \fi ##1}}%
\ps@@fancy
\gdef\ps@fancy{\@fancyplainfalse\ps@@fancy}%
\ifdim\headwidth<0sp
\global\advance\headwidth123456789sp\global\advance\headwidth\textwidth
\fi
}
\makeatother

\makeindex

\begin{document}
\baselineskip=16.0pt
\newcounter{Mycounter}\setcounter{Mycounter}{\value{chapter}}
\addtocounter{chapter}{8}
\chapter{
Mixed phases during the phase transitions
}
\label{chap:9}

\vspace*{-0.5cm}
\centerline{
Toshitaka Tatsumi{\footnote{E-mail: tatsumi@ruby.scphys.kyoto-u.ac.jp}}, 
Nobutoshi Yasutake{\footnote{E-mail: yasutake@resceu.s.u-tokyo.ac.jp}}, 
and Toshiki Maruyama{\footnote{E-mail: maruyama.toshiki@jaea.go.jp}
}}
\centerline{
$^{1}$Department of Physics, Kyoto University, Kyoto 606-8502, Japan
}
\centerline{$^2$Research Center for the Early Universe, School of Science, University of Tokyo,}
\centerline{Bunkyo-ku, Tokyo 113-0033, Japan}

\centerline{
$^3$Advanced Science Research Center,
Japan Atomic Energy Agency,}
\centerline{
Shirakata Shirane 2-4, Tokai, Ibaraki 319-1195, Japan
}

\

Quest for a new form of matter inside compact stars compels us to examine the thermodynamical properties of the phase transitions. 
We closely consider the first-order phase transitions and the phase equilibrium on the basis of the Gibbs conditions, 
taking the liquid-gas phase transition in asymmetric nuclear matter as an example. 
Characteristic features of the mixed phase are figured out 
by solving the coupled equations for mean-fields and densities of constituent particles self-consistently within the Thomas-Fermi approximation. 
The mixed phase is inhomogeneous matter composed of two phases in equilibrium; it takes a crystalline structure with a unit of various geometrical shapes, 
inside of which one phase with a characteristic shape, called ``pasta'', is embedded in another phase by some volume fraction. 
This framework enables us to properly take into account the Coulomb interaction and the interface energy, 
and thereby sometimes we see the mechanical instability of the geometric structures of the mixed phase.  
Thermal effect on the liquid-gas phase transition is also elucidated.         

Similarly hadron-quark deconfinement transition is studied in hyperonic matter, 
where the neutrino-trapping effect as well as the thermal effect is discussed in relation to the properties of the mixed phase.   
Specific features of the mixed phase are elucidated and the equation of state is presented.

\section{Introduction}

Equation of state (EOS) for matter inside compact stars is sometimes characterized by the phase transitions.
There have been expected various first-order phase transitions in nuclear matter, such as liquid-gas transition in the low densities, meson condensations and hadron-quark deconfinement transitions at high-density region \cite{sha,tama,glen,hae}. If such phase transitions occurs inside compact stars, they affect EOS as well as thermal, dynamical or magnetic properties \cite{bec,sag,wata09,sot}. There has been a theoretical development and recent studies have shown that the inhomogeneous states called ``pasta'' may emerge as a mixed phase during the phase transitions, based on the Gibbs conditions \cite{revpasta}. 
Historically, the mixed phases have been mostly treated very naively by applying the Maxwell construction to get the EOS in thermodynamic equilibrium. Ravenhall et al.\ discussed the possible emergence of the pasta structures at low-density nuclear matter, since uniform nuclear matter is not stable at low densities \cite{Rav83}. Their pasta phase may be regarded as a mixed phase following liquid-gas phase transition. 
In 1990 Glendenning has pointed out that the phase equilibrium in multi-component system or in the case of plural chemical potential corresponding to the conserved quantities such as in neutron star matter must be carefully treated based on the more general Gibbs conditions \cite{glen, gle92}. He emphasized that the phase transitions should be treated in a proper way in 
neutron-star matter, which is specified by two conserved quantities, electromagnetic charge as well as baryon number. In particular the local charge-neutrality condition used in the case of the Maxwell construction should be replaced by the global one in the mixed phase.
Consequently the pressure becomes no more constant as in the Maxwell construction, but still increases in the mixed phase. He also demonstrated that the mixed phase is extended in the rather wide density region within a bulk calculation, where two sets of semi-infinite matter are considered in phase equilibrium, separated by a sharp boundary, and the pressure balance, chemical equilibrium and the global charge-neutrality conditions are imposed in the absence of the Coulomb interaction. It is important to see that particle fractions in each phase are no more the same to each other for given particle fractions as a whole. Nowadays we know that such phase transitions are rather common in other fields as well \cite{mag96,ron,khra06,for07,net08,ios} and we call them {\it non-congruent} phase transitions, following Iosilevskiy \cite{ios}. Generally any phase transition in a system with two or more conserved charges must be non-congruent.

In this calculation the Coulomb interaction is discarded and we can by no means estimate how important it is, or the surface energy can not be taken into account. If we consider  a more realistic situation, we must consider inhomogeneous matter with various geometric structures called pasta, and take into account the finite-size effects such as the Coulomb interaction and the surface tension. Heiselberg et al.\ have demonstrated that such finite-size effects are important for the mixed phase in the context of hadron-quark transition \cite{HPS93}. Treating the pasta structure as in ref.~\cite{Rav83}, they simply considered uniform particle densities in both phases. They showed a possibility of that the finite-size effects may largely restrict the pasta region.
Subsequently, Voskresensky et al.\ have shown that the rearrangement of particle densities and the screening effect for the Coulomb interaction becomes important, and  the large surface tension gives the mechanical instability of the geometrical structures in the mixed phase \cite{voskre}. Following this idea, we have studied the pasta structures brought by the kaon condensation and hadron-quark deconfinement transition \cite{marukaon06,end05,hyp07}. More recently we have extended our framework to deal with the finite temperature case or the neutrino-trapping matter, which may be relevant to supernova explosions or neutron star mergers \cite{yas09}. 

In this chapter we first present thermodynamic concepts of the non-congruent transition by considering the liquid-gas phase transition in asymmetric nuclear matter (section \ref{cong}). In section \ref{secLG} some results will be presented by a numerical calculation about the pasta structures in asymmetric nuclear matter. In section \ref{secHQ} we will present the results about hadron-quark deconfinement transition, where the effects of finite temperature or neutrino trapping are also discussed. Section \ref{sum} is devoted to summary and concluding remarks.

\section{Phase equilibria in thermodynamics}
\label{cong}

\subsection{Maxwell construction in the pure system}

\begin{figure}[h]
\begin{minipage}{0.56\textwidth}
\begin{center}
\includegraphics[width=6cm]{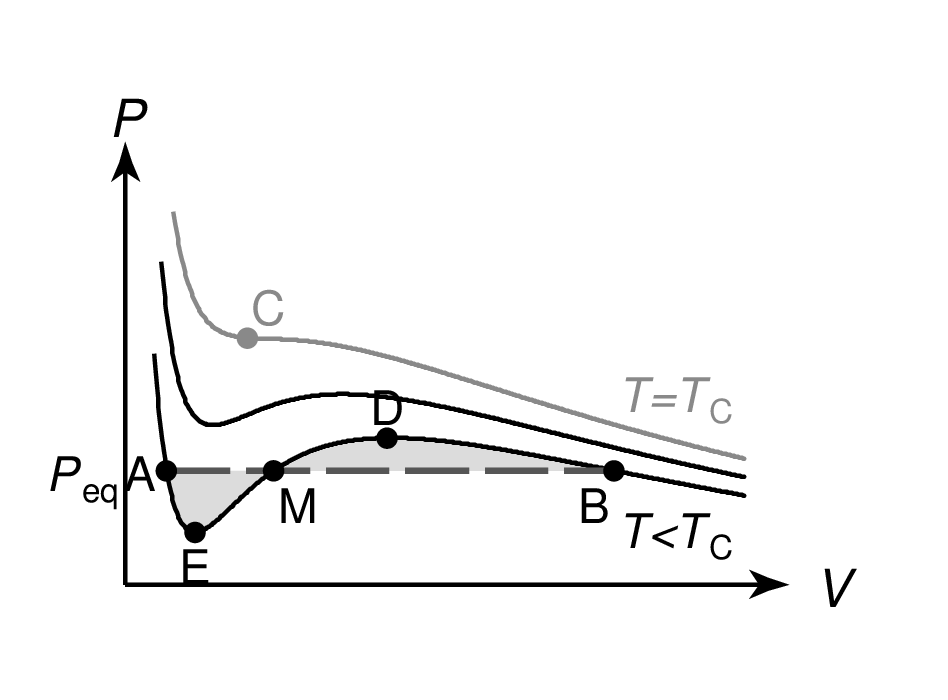}
\end{center}
\caption{Isotherms of the van der Waals fluid on the $P-V$ plane.}
\label{maxwell}
\end{minipage}
\hspace{\fill}
\begin{minipage}{0.4\textwidth}
\begin{center}
\includegraphics[width=6cm]{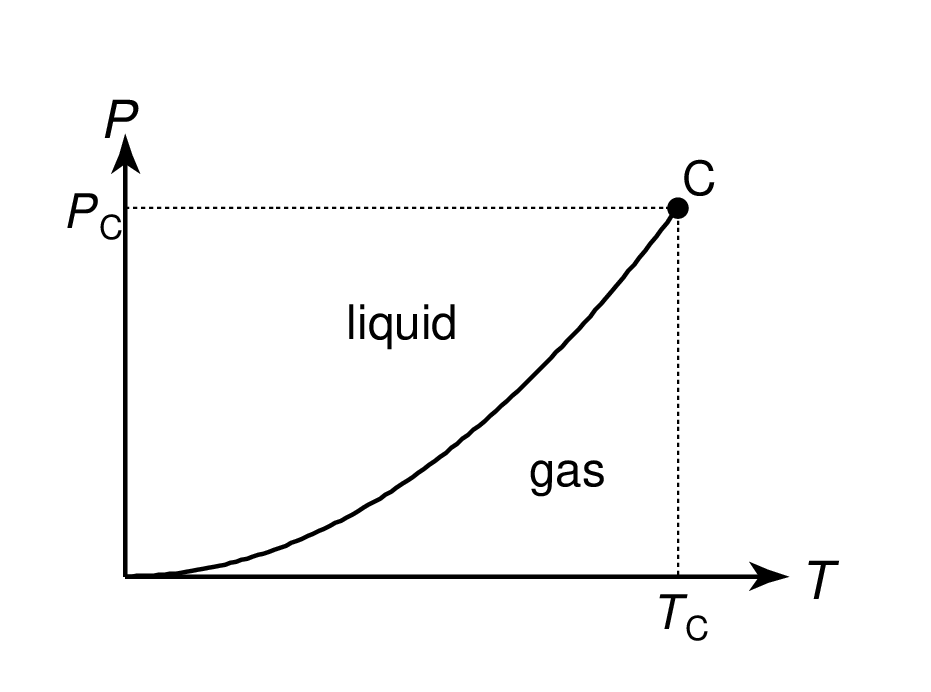}
\end{center}
\caption{Saturation curve on the $P$-$T$ plane.}
\label{saturation}
\end{minipage}
\end{figure}

Before beginning our discussion about the non-congruent transitions, let us briefly summarize more familiar case in the pure system, where the Maxwell rule can be applied to construct the equilibrium curve. Since there is only a single component, it is trivially congruent.  Typical isotherms for liquid-gas phase transition are sketched in Fig.\ \ref{maxwell}, where we can see that the gas phase (low-density phase) and the liquid phase (high-density phase), both of which are thermodynamically stable due to $\partial P/\partial V<0$ are clearly separated by the thermodynamically unstable region $\partial P/\partial V>0$ at temperature $T$ below the critical temperature $T_c$. 
Above $T_c$ two phases are indistinguishable\footnote{
Recently the supercritical fluid above the critical point has attracted much interest, especially in the context of various applications \cite{tester}.}.
The Gibbs conditions for phase equilibrium read in this case as $P^G=P^L, \mu^G=\mu^L$ besides $T_G=T_L$ in terms of pressure $P_i$, chemical potential $\mu_i$ and temperature $T_i$ in each phase $i=G$ or $L$. We can then construct the EOS by the graphical method, the Maxwell construction (see Fig.\ \ref{maxwell}). 

For a given $T$ and fixed number, points $A$ and $B$ denote the initial and final points of the phase coexistence, 
subjected to the equal-area condition, 
\beq
\int_{AEM}V(P)dP=\int_{BDM}V(P)dP,
\eeq 
which is equivalent with $\mu^L=\mu^G$ by way of the thermodynamic relation, 
$dG=Nd\mu=VdP$ with the Gibbs free energy $G$. 
Alternatively the double tangent method can be applied for the Helmholtz free energy $F=G-PV$. 
Anyway we can draw the {\it binodal curve} (coexistence curve) by connecting the equilibrium points, $A$ and $B$, by changing $T$. 
At the same time we can also draw the {\it spinodal curve} by connecting the limit points of the stable region $D$ and $E$ 
where $\partial P/\partial V=0$ by changing $T$. 
At the point $E$, where $\partial^2P/\partial V^2>0$, the phase is intrinsically stable against the shrinkage of the volume, 
while it is unstable for the expansion of the volume.  
The similar argument can apply for the point $D$ by exchanging the volume variations. 
Consequently, the segments $AE$ and $BD$ show the metastable regions.

We can draw the saturation curve, which is defined by the equilibrium pressure as a function of $T$, 
on the pressure-temperature plane, as in Fig.\ \ref{saturation}. 
Then two phases are separated by a single curve terminated at the critical temperature $T_c$, 
which means $P$ is constant in the mixed phase. 
In the following section we shall see that it spreads to involve a finite mixed-phase region 
in the general case with multi-component, since pressure still increases in the mixed phase. 
More interestingly the mixed phase emerges with peculiar geometrical structures.  

\subsection{Non-congruent transition and structured mixed phase within the bulk calculation}
First we briefly summarize the basic concepts of the non-congruent transition in a clear fashion by the bulk calculation without the Coulomb interaction or surface energy. We consider the liquid-gas transition at subnuclear densities in asymmetric nuclear matter (ASNM) as an example \cite{chom04}. The proton-number fraction $Y_p\equiv N_p/N$ then specifies ASNM.

It is well-known that EOS of nuclear matter exhibits a thermodynamically forbidden region, which separates ``liquid'' and ``gas'' phases  at subnuclear densities. 
This phase transition looks like the first-order, so that we must 
treat the phase-equilibrium in the mixed phase.
Since there are two conserved charges, electromagnetic charge and baryon number  in ASNM, we must discuss the phase transition 
in the {\it binary} system. 
Different from the pure component system, thermodynamic behavior of the mixed phase becomes complicated in this case  \cite{tester,lan}; two coexisting phases can vary the chemical composition with keeping the total charges.

Thermodynamical potential for ASNM is defined as 
\beq
\Omega(T,V,\{\mu_i\}_{i=n,p})=-\beta^{-1}{\ln}\;{\rm Tr}\;{\exp}\left[-\beta\left({\hat H}-\sum_{i=n,p}\mu_i {\hat N}_i\right)\right].
\eeq
Then particle number and pressure are given as
\beqa
N_i(T,V,\{\mu_j\}_{n,p})=-\left(\frac{\partial\Omega}{\partial\mu_j}\right)_{T,V,\mu_{j\neq i}}\nonumber\\
P(T,V,\{\mu_i\}_{n.p})=-\left(\frac{\partial\Omega}{\partial V}\right)_{T,\{\mu_i\}_{n,p}}.
\label{np}
\eeqa 
The Helmholtz free energy can be expressed as 
\beqa
F(V,T,N_p,N_n)&=&V{\cal F}(T,\{\rho_i\}_{n,p})\nonumber\\
&=&\Omega(T,V,\{\mu_i\}_{n,p})+V\sum_{i=n,p}\mu_i\rho_i(T, \{\mu_i\}_{n,p}),
\eeqa
with the particle-number density $\rho_i$ ,
\beq
N_i=V\rho_i(T,\{\mu_j\}_{n,p})
\eeq
for uniform matter. The Gibbs-Duhem relation gives $\Omega=-PV$.

Usually EOS ($P$ vs $V$) renders the Van der Waals type and thermodynamically unstable region ($dP/dV<0$) appears at subnuclear density, which signals phase transition (called the ``liquid-gas'' phase transition in nuclear matter) \cite{chom04}. It looks like first-order phase transition, so that there appears the mixed phase composing of two phases in equilibrium. 
Since there are two conserved charges, baryon number ($N_B=N_p+N_n=V\rho_B$) and isospin number ($I_3=N_p-N_n=V\rho_I$), we must carefully treat the phase equilibrium in the mixed phase of this binary system\footnote{
In neutron-star matter electric charge is conserved instead of isospin at low temperature. Moreover, lepton number is also conserved in some situations as in supernovae, which makes a ternary system \cite{pag}.}. 

In the following we assume the equilibrium between two phases (liquid(L) and gas(G)) in the mixed phase. Then the Gibbs conditions read
\beq
T^G=T^L,\quad P^G=P^L,\quad \mu_i^G=\mu_i^L (i=n,p).
\label{gibbs}
\eeq 

The phase diagram is then given in the $T$-$P$-$Y_p$ space: phase-separation boundary is then given as a two dimensional surface.
Note that the familiar Maxwell construction cannot be applied in this binary system, since it works only for single component system. When we draw the pressure surface as a function of $\mu_p$ and $\mu_n$ for given $Y_p$ and $T$. we can graphically see a folded surface and the intersection gives the coexisting curve. The mixed phase develops along the coexisting curve and obviously satisfies the Gibbs conditions. As a result, pressure is no more constant in the mixed phase. So this is not first-order phase transition, but second-order one in the sense of Ehrenfest \cite{stan}. However, since it never accompanies the change of any symmetry, it is first-order phase transition in the sense of Landau \cite{lan}. We shall see that the mixed phase consists of two distinct phases in equilibrium and there exist a metastable state and latent heat, which should characterize the first-order phase transition. 

When we plot the beginning and terminating pressures in the $P$-$T$ plane by changing $T$ for a given $Y_p$, we have an extended coexisting domain. The phase-separation boundary is called {\it binodal curve}. Following Iosilevskiy, we call it {\it non-congruent} phase transition. Note that $\rho_I$ as well as $\rho_B$ takes the different value in each phase, keeping the volume average fixed. 
It is instructive to seek the coexisting region by daring to apply the Maxwell construction, as in the van der Waals fluid (Fig.~\ref{maxwell}). One may see only a curve on the same plane, which is called {\it congruent} phase transition and the water-vapor phase transition is a typical example \cite{ios}. When we apply the Maxwell construction to the ASNM, the first two conditions in (\ref{gibbs}) are obviously satisfied. However, the chemical equilibrium is partially violated since the proton fraction is forced to be the same in both phases: actually the baryon chemical potential, $\mu_B=\mu_p+\mu_n$, results in being equal at the end points, while the isospin chemical potential, $\mu_I=\mu_p-\mu_n$, is different.

Consider the manifold $\mathcal F$ in the parameter space spanned by $\{\rho_i\}_{n,p}$ with $T$ fixed. 
If there is a common tangent plane connecting two distinct configurations with $\{\rho_i'\}_{n,p}$ in G and $\{\rho_i''\}_{n,p}$ in L, Gibbs' criteria are satisfied. 
Then the mixed phase can be constructed by combining two phases with volume fraction $\lambda=V^L/V(=V^L+V^G), 0<\lambda<1$. 
Total charge densities are given by $\rho_i=\lambda\rho''_i+(1-\lambda)\rho'_i$ and Helmholtz free energy can be written as 
\beq
{\cal F}(T,\{\rho_i\}_{n,p})=\lambda{\cal F}(T,\{\rho''_i\}_{n,p})+(1-\lambda){\cal F}(T,\{\rho'_i\}_{n,p}).
\eeq

\begin{figure}[h]
\begin{center}
\includegraphics[width=0.4\textwidth]{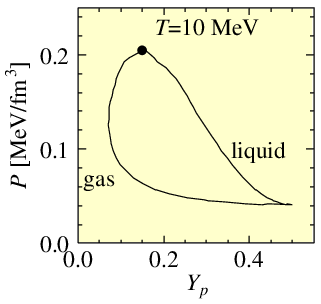}
\end{center}
\caption{Binodal curve for ASNM at $T=10$MeV (see section \ref{secBulk}). 
The filled circle denotes the critical point.}
\label{binodal}
\end{figure}

The binodal points for given $Y_p$ are then obtained by considering the two limits, $\lambda=0,1$
\footnote{Note that there is an exception: we shall see a possibility of the existence of $\lambda_{\rm max}$ in the special case, called retrograde condensation.}
. From Eq.~(\ref{np}) ,
\beq
P^G(\mu_i)=P^L(\mu_i),
\label{peq}
\eeq
and
\beq
\frac{\partial P^G}{\partial\mu_i}=\rho_i
\label{pdeq1}
\eeq
for $\lambda=0$, and
\beq
\frac{\partial P^L}{\partial\mu_i}=\rho_i
\label{pdeq2}
\eeq
for $\lambda=1$. Three equations in (\ref{gibbs}) or Eqs.~(\ref{peq}),(\ref{pdeq1}),(\ref{pdeq2}) determine the unknown values of chemical potentials $\mu_i$ and the baryon densities for the binodal points. Finally the binodal curve is obtained by connecting the binodal points by changing the mixture $Y_p$ (see Fig.~\ref{binodal}).
Note that $P^L(\mu_i)$ in the former case and $P^G(\mu_i)$ in the latter case never correspond to the $\lambda=0$ 
and $\lambda=1$, respectively.

\subsection{Stability criteria}

\begin{figure}[h]
\begin{minipage}{0.4\textwidth}
\begin{center}
\includegraphics[width=6cm]{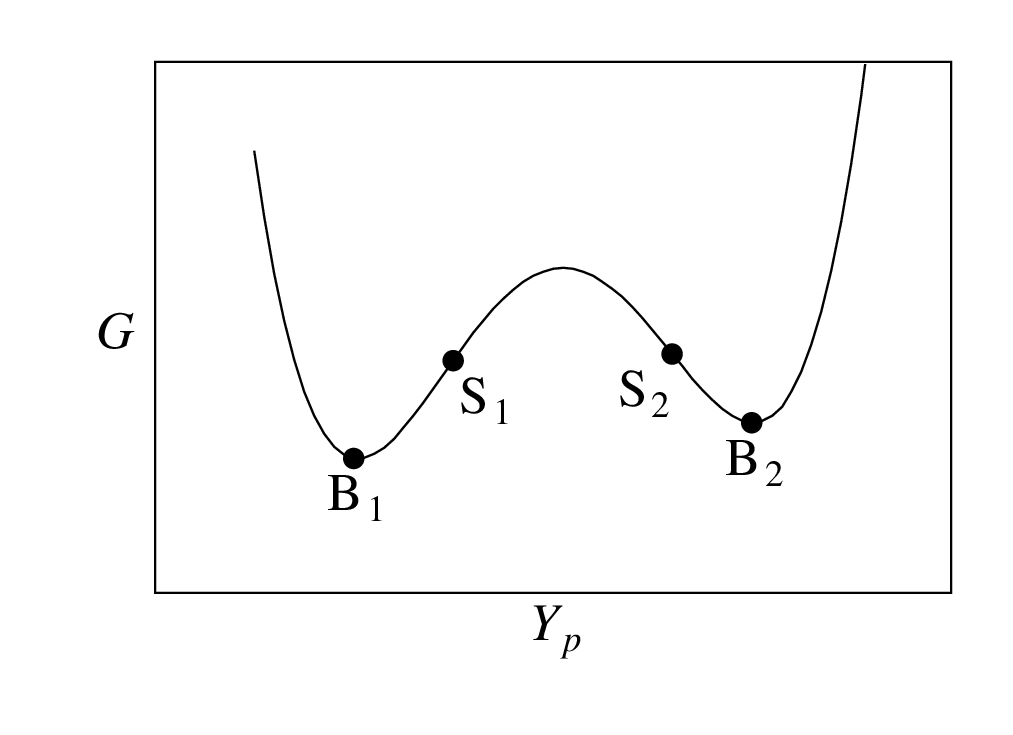}
\end{center}
\caption{Schematic view of the Gibbs free energy $G$ at fixed temperature as a function of $Y_p$. 
$B_{1.2}$ denote the binodal points and $S_{1,2}$ the spinodal points where $\left(\partial^2 G/\partial Y_p^2\right)_{T,P}=N_B(\partial\mu_p/\partial Y_p)_{T,P}=0$.}
\label{gibbs}
\end{minipage}
\hspace{\fill}
\begin{minipage}{0.56\textwidth}
\begin{center}
\includegraphics[width=6cm]{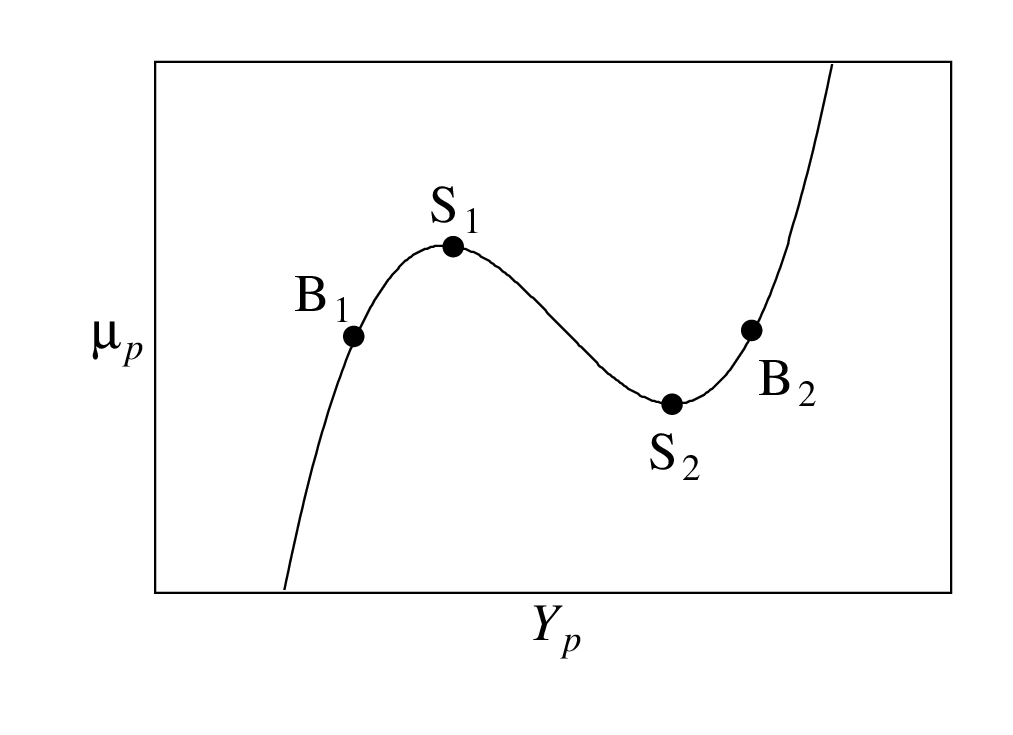}
\end{center}
\caption{Proton chemical-potential isobar at fixed temperature as a function of $Y_p$. 
The segment between $S_{1,2}$ denotes the spinodal region, where $\left(\partial\mu_p/\partial Y_p\right)_{T,P}<0$, 
and the segments between $B_{1,2}$ and $S_{1,2}$ correspond to the metastable regions.}
\label{spinodal}
\end{minipage}
\end{figure}

{\it Spinodal} instability has been often discussed in ASNM in relation to the nuclear multi-fragmentation in heavy-ion collisions or in the crust region of neutron stars \cite{chom04, pet95,muller95,marg03,prov06,rand09}. Since it should be also related to the formation of the pasta structures, we briefly summarize it. The stability criterion implies the convexity of the free energy density $\mathcal F$ of $\rho_i$: Hessian or the curvature matrix
\beq
C=\left(\begin{array}{cc}
\partial^2{\mathcal F}/\partial\rho_n^2&\partial^2{\mathcal F}/\partial\rho_n\partial\rho_p\\
\partial^2{\mathcal F}/\partial\rho_p\partial\rho_n&\partial^2{\mathcal F}/\partial\rho_p^2
\end{array}\right)
=\left(\begin{array}{cc}
\partial\mu_n/\partial\rho_n&\partial\mu_n/\partial\rho_p\\
\partial\mu_p/\partial\rho_n&\partial\mu_p/\partial\rho_p
\end{array}\right)
\label{curv}
\eeq
has only positive eigenvalues\footnote{
In neutron-star matter, which is charge-neutral in the presence of electrons, the curvature matrix is modified by adding 
$\partial\mu_e/\partial\rho_e$ to $\partial\mu_p/\partial\rho_p$ \cite{pet95,prov06}.},
\beq
{\rm tr}(C)\geq 0 ~~~{\rm and}~~~{\rm det}(C)\geq 0.
\eeq
Since both eigenvalues never become negative in ASNM at subnuclear densities, 
one negative eigenvalue or equivalently ${\rm det}(C)<0$ defines the spinodal region. 
The condition ${\rm det}(C)<0$  is further equivalent with 
\beq
\left(\frac{\partial P}{\partial\rho_B}\right)_{T,Y_p}<0 ~~~{\rm or}~~~\left(\frac{\partial\mu_p}{\partial Y_p}\right)_{T,P}<0,
\eeq
where the first inequality is called ``mechanical'' instability and the second one ``diffusive(chemical)'' instability \cite{muller95}.
The mechanical instability is the familiar one and comes from the negative compressibility, and the spinodal condition is given only by this inequality in symmetric nuclear matter (SNM). However, the chemical instability first happens in ASNM, whereas curvature always remains positive and no spinodal instability is predicted in pure neutron matter \cite{marg03}. So there should be a lower bound for $Y_p$, below which ASNM is always stable. The Gibbs free energy $G$ is depicted in Fig.~\ref{gibbs}, whose derivative with respect to proton number gives proton chemical potential, $(\partial G/\partial N_p)_{T,P}=\mu_p$. In Fig.~\ref{spinodal} chemical potential isobar is shown  for protons. 
In the region between the binodal points $B_{1,2}$ the uniform matter is unstable and separated to two phases with different $Y_p, (0\leq Y_p\leq 0.5)$
\footnote{The region $0.5\leq Y_p\leq 1$ is redundant due to symmetry of interchanging protons and neutrons. }
; the spinodal region is specified by the spinodal points $S_{1,2}$ and two regions between $B_1$ and $S_1$ and $B_2$ and $S_2$ are the metastable regions. As ASNM enters the spinodal region, a unique mechanism of phase separation works and small fluctuations of $Y_p$ grow to produce a non-uniform random pattern. This phenomenon is known as {\it spinodal decomposition} \cite{tester}.  In the metastable regions ASNM is stable for small fluctuations of $Y_p$, but more likely becomes inhomogeneous by way of {\it nucleation} and
 growth. These features should be peculiar to the first-order phase transitions.

\subsection{Finite-size effects and ``Pasta'' structures}

We have only discussed the binary system to demonstrate the characteristic features of the non-congruent transition. Since there are usually two conserved quantities, baryon number and 
electromagnetic charge 
in the cold catalyzed matter, many cases are treated as the binary systems inside compact stars.
Extension of the discussion to multi-component system is straightforward but very complicated. In the neutrino trapping case as in supernova explosions, we need another conserved quantity, lepton number. Thus we must consider the ternary matter. In the recent paper this problem has been generally addressed and classification of various situations has been presented \cite{pag}. 

\begin{figure}[h]
\begin{center}
\includegraphics[width=0.9\textwidth]{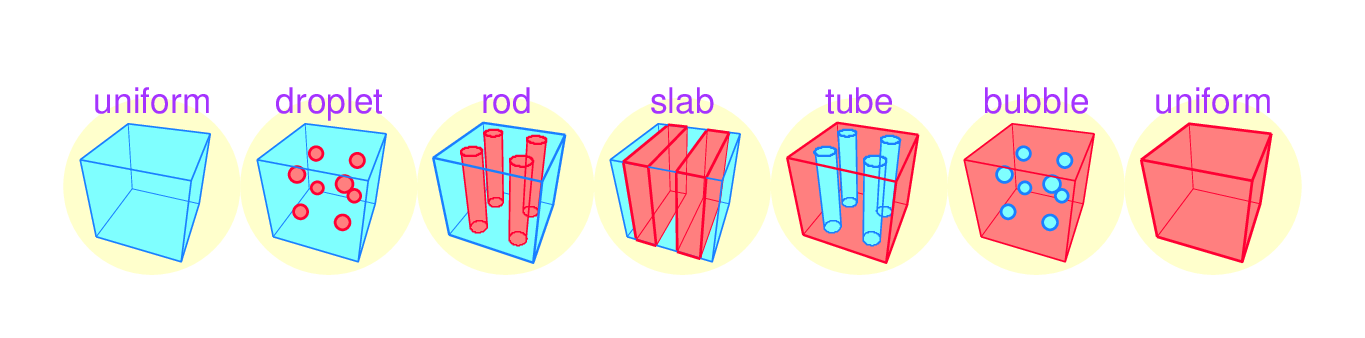}
\end{center}
\caption{Schematic picture of ``pasta'' structures.
When red phase appears in the blue phase, the red phase
form droplets in the sea of blue.
With increase of the volume fraction of red phase in the blue phase,
the shape of the red phase changes from droplet to rod, 
slab, tube, and to bubble before pure red phase appears.}
\label{figSchematicPasta}
\end{figure}

We have presented general features of the non-congruent transition in a clear fashion with a help of the bulk calculation, discarding the Coulomb interaction and geometrical structures which are called the finite-size effects. However, the bulk calculation is insufficient to describe the essential aspects of the non-congruent transition. 
We can see the emergence of various geometrical shapes, called pasta (see Fig.~\ref{figSchematicPasta}), by the cooperative effect of the Coulomb interaction and the surface energies, and the importance of the charge screening effect which is a typical many-body effect and makes pasta structures unstable, called mechanical instability.

\begin{figure}[h]
\begin{center}
\includegraphics[width=0.6\textwidth]{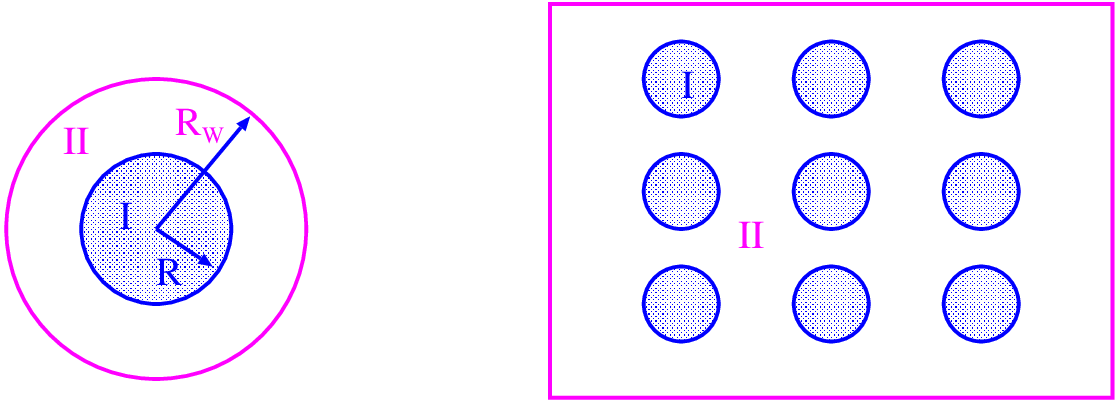}
\caption{Wigner-Seitz cell for the droplet case (left panel) and crystalline structure composed of droplets (right panel).}
\label{ws}
\end{center}
\end{figure}
\begin{figure}[h]
\begin{center}
\includegraphics[width=0.4\textwidth]{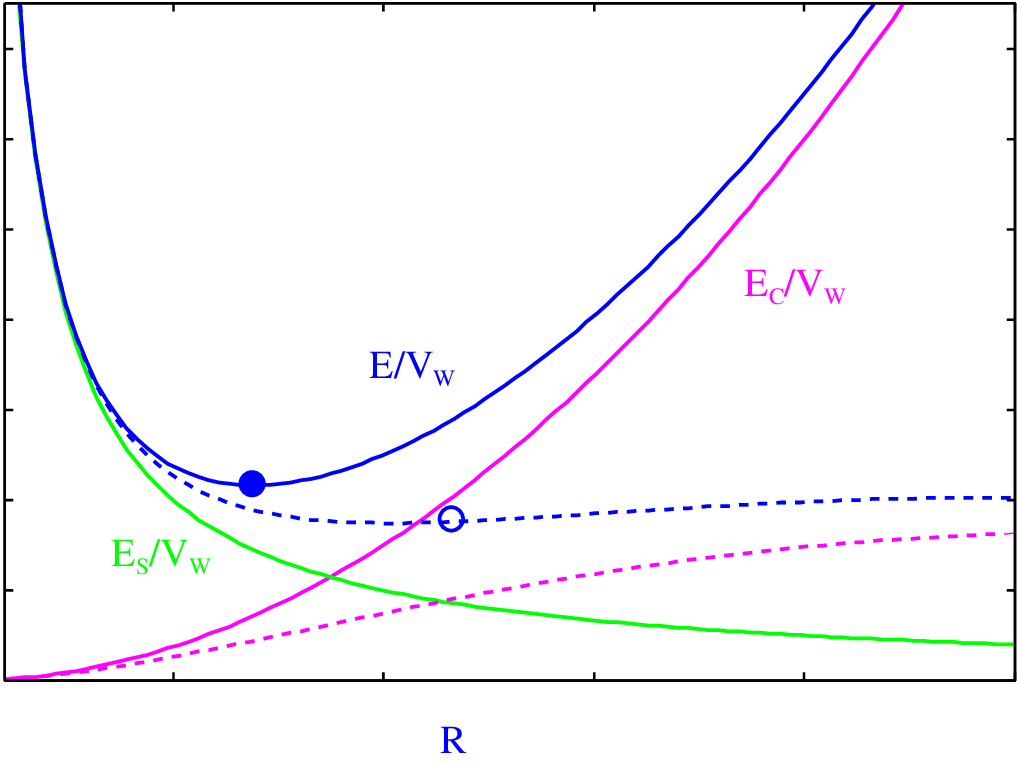}
\caption{Energy densities as functions of the radius of droplets. Solid curves show the case without the screening effect, while dotted curves with screening.}
\label{bulk}
\end{center}
\end{figure}

A rough estimate about the finite-size effects can be done as a straight extension of the bulk calculation. Since the crystalline structure appears in this case we use the Wigner-Seitz cell approximation (see Fig.\ \ref{ws}). Consider a simple model with a droplet composed of the phase I embedded in the phase II in a Wigner-Seitz cell with the volume $V_W=4\pi R^3/3$ at $T=0$: particle density is assumed to be uniform in each phase, charge neutrality is achieved by electrons and  
their interface is represented by the sharp surface with a surface tension parameter $\tsurf$.  For a given volume fraction $\lambda=(R/R_W)^3$, the total energy then consists of three parts,
\beq
E=E_V+E_C+E_S,
\label{etot}
\eeq
where $E_V$ is the volume energy computed with the strong-interaction Hamiltonian, and $E_C$ and $E_S$ are the Coulomb energy and the surface energy, respectively.
The energy density $E_V/V_W$ does not depend on $R$, while the Coulomb-energy density renders
\beq
E_C/V_W=\lambda\frac{16\pi^2}{15}(\rho_{\rm ch}^I-\rho_{\rm ch}^{II})^2R^2,
\label{ecoul}
\eeq 
and the surface-energy density 
\beq
E_S/V_W=\lambda\frac{3\tsurf}{R}.
\label{esurf}
\eeq
In Fig.\ \ref{bulk} we sketch the $R$ dependence of these energy densities. The optimal value of $R$ is then given by the minimum condition,
\beq
\frac{\partial (E/V_W)}{\partial R}=0.
\eeq
Using Eqs.~(\ref{etot}), (\ref{ecoul}), (\ref{esurf}), we can see the well-known relation $E_S=2E_C$ holds at the minimum. 

We shall see this simple consideration is too crude to describe the mixed phase in a realistic way. Once the Coulomb interaction is included, 
we must take into account the rearrangement of the particle densities at the same time, 
and in turn the screening effect for the Coulomb interaction. Consequently the Coulomb interaction becomes effectively short-range by the Debye mass,
$m_D^2=\sum_i 4\pi\partial\rho_{\rm ch}^i/\partial\mu_i$. As shown in Fig.\ \ref{bulk} the Coulomb-energy density is no more an increasing function with respect to $R$, 
but is exponentially reduced at large $R$. The minimum point is shifted to larger $R$ as $\tsurf$ increases, 
which gives rise to a mechanical instability of the droplet above the critical value of $\tsurf$. The mechanical instability of the pasta structure also occurs in other geometrical shapes, which means that both $R$ and $R_W$ go to infinity with their ratio fixed. Thus we recover the picture of the phase equilibrium of two bulk matters, where the surface energy is irrelevant and the Coulomb energy vanishes due to the achievement of the local charge-neutrality. These features indicate that the EOS resembles the one given by the Maxwell construction.

The rearrangement of the particle densities also induces a change in the volume energy $E_V$, which is composed of kinetic energy and strong-interaction energy. In the above calculation $E_V$ itself is intrinsically large but independent on $R$, since we have assumed uniform densities $\rho_i^{I,II}$ in each phase. Once the Coulomb interaction is properly taken into account, particle densities are no more uniform to produce a $R$ dependence in $E_V$. The difference in $E_V$ is called the correlation energy in ref. \cite{voskre}, defined by 
\beq
E_{\rm corr}\equiv E_V(\rho_i({\bf r}))-E_V(\rho_i^{I,II}).
\eeq
The detailed analysis showed that $E_{\rm corr}$ gives negative values and behaves $R^{-1}$ for large $R$. Sometimes $E_{\rm corr}$ gives a significant contribution as later seen in hadron-quark deconfinement transition (section \ref{secHQ}). 

\section{Liquid-gas transition in asymmetric nuclear matter}\label{secLG}

The properties of inhomogeneous low-density matter was
already studied in the early 70th by Baym et al.\ \cite{Baym71}.
They have presented a concept of structured mixed phase which
includes nuclear droplet in the neutron gas as well as its
reversal structure (droplet of gas in the nuclear sea).

About a decade later, Ravenhall et al.\ \cite{Rav83} have considered 
not only the three-dimensional structures but also those of
one and two dimension. 
They have introduced a dimensionality $d$ with a fractional number
as well as other parameters such as volume fraction, density in
the nuclear phase, and so on.
With given average density and proton fraction, they looked for
the parameters which give a minimum free energy.
With increasing the average density, 
the structure of nucleus changes from 
droplet to rod, slab, tube, bubble, and finally uniform.
Afterward, such a series of structures was called ``pasta'' (see Fig.\ \ref{figSchematicPasta}).
The corresponding $d$ changes from 3 for droplet to 2 for rod,
1 for slab, 2 for tube, and 3 for bubble.
But $d$ changes smoothly, which suggests the existence
of intermediate shapes besides the typical pasta structures.

The next year, Hashimoto et al.\ \cite{Has84} used a liquid-drop model
with the Wigner-Seitz (WS) cell approximation,
and showed that the stable nuclear shape changes
as in the same way as has shown by the above Ref.\ \cite{Rav83}.
After that there appeared studies of pasta structures by several groups
using different frameworks which are, however, basically mean-field models \cite{Oya93,Lor93,Cheng97,she98,goge,nak}.

Besides the mean-field approaches,
molecular dynamics (MD) simulations \cite{Aic91,Mar98} 
were applied to the inhomogeneous
low-density matter \cite{wata09,Mar98, Gen00,wata04}. 
The advantage of MD simulation is that one needs to
assume no structure, or capability in simulating the dynamical processes \cite{wata09}. 
Some mean-field approaches are also possible without the assumption of
the structure \cite{Wil85, oka}.

\subsection{Framework based on relativistic mean-field theory}

Here we use the relativistic mean-field (RMF) model to describe asymmetric nuclear matter.
Exploiting the idea of the density functional theory (DFT) within RMF model \cite{refDFT}, 
we can calculate the particle-density profiles as well as thermodynamic quantities.
The Lagrangian density of the system is written as
\begin{eqnarray}
{\cal L}&=&{\cal L}_N+{\cal L}_M+{\cal L}_e \\
{\cal L}_N&=& \bar \Psi\left[ i\gamma ^\mu \partial_\mu -m^*_N-g_{\omega N}\gamma^\mu \omega_\mu 
    - g_{\rho N}\gamma^\mu {\vec \tau}{\vec b}_\mu -e\frac{1+\tau_3}{2} \gamma^\mu V_\mu\right ]\Psi\\ 
{\cal L}_M&=&+\frac{1}{2}(\partial_\mu\sigma)^2-\frac{1}{2}m_\sigma^2\sigma^2-U(\sigma)  \nonumber\\
           &&-\frac{1}{4}\omega_{\mu\nu}\omega^{\mu\nu}+\frac{1}{2}m_\omega^2\omega_\mu\omega^\mu  \nonumber\\
           &&-\frac{1}{4}{\vec R}_{\mu\nu}{\vec R}^{\mu\nu}+\frac{1}{2}m_\rho^2{\vec R}_\mu{\vec R}^\mu \\ 
{\cal L}_e&=&-\frac{1}{4}V_{\mu\nu}V^{\mu\nu}
     + \bar\Psi_e\left[i\gamma^\mu\partial_\mu-m_e+e\gamma^\mu V_\mu\right]\Psi_e,\\
m_N^*&=&m_N-g_{\sigma N}\sigma\\
U(\sigma)&=&\frac{1}{3}bm_N(g_{\sigma N}\sigma)^3-\frac{1}{4}c(g_{\sigma N}\sigma)^4
\end{eqnarray}
where $F_{\mu\nu}$ stands for $\partial_\mu F_\nu-\partial_\nu F_\mu$.
By the Euler-Lagrange equation
\beq
\partial_\mu\left[\frac{\partial {\cal L}}{\partial(\partial_\mu\phi)}\right]-\frac{\partial{\cal L}}{\partial\phi}=0,~~~(\phi=\sigma,\omega_\mu,R_\mu,V_\mu, \Psi)
\eeq
we get the coupled equations of motion for the scalar ($\sigma$), isoscalar-vector 
($\omega$), isovector-vector ($\rho$),
electromagnetic and nucleon fields.
Since the spatial components of vector fields become zero due to the isotropy of the system,
the meson mean-fields ($\sigma, \omega_0, R_0$) and electric field ($V_0$) obey the following equations of motion
\beqa
-\nabla^2\sigma({\bf r})+m_\sigma^2\sigma({\bf r}) &=& -{dU\over d\sigma}({\bf r})+g_{\sigma N}(\rho_n^{(s)}({\bf r})+\rho_p^{(s)}({\bf r}))  \label{eq:meson1}\\
-\nabla^2\omega_0({\bf r})+m_\omega^2\omega_0({\bf r}) &=& g_{\omega N}(\rho_p({\bf r})+\rho_n({\bf r})) \\
-\nabla^2R_0({\bf r})+m_\rho^2R_0({\bf r}) &=& g_{\rho N}(\rho_p({\bf r})-\rho_n({\bf r}))  \\
\nabla^2{V_0({\bf r})} &=& 4\pi e^2{\rho_{\rm ch}({\bf r})} \ \ \ \ 
(
{\rho_{\rm ch}({\bf r})}\;=\;{\rho_p({\bf r})}-{\rho_e({\bf r})}) .\label{eq:meson4}
\eeqa
For nucleons and electrons we employ the Thomas-Fermi approximation at finite temperature $T$ and 
the distribution function $f$ of proton, neutron and electron can be written as
\beqa
f_{i=n,p}({\bf r};{\bf p}, \mu_i)&=&\left(1+\exp\left[\left(\sqrt{p^2+{m_N^*({\bf r})}^2}-\nu_{i}({\bf r})\right)/T\right]\right)^{-1},\\
f_e({\bf r}; {\bf p}, \mu_e)&=& \left(1+\exp\left[\left(p-(\mu_e-V_0({\bf r}))\right)/T\right]\right)^{-1}.\\
\eeqa
For simplicity we neglect contributions of anti-particles,
which is legitimate for high chemical potentials $\mu_i$ and low $T$.
Using above distribution functions $f_a, (a=n,p,e)$, relations between chemical potentials $\mu_a$ and densities $\rho_a$ are written as
\beqa
\rho_e({\bf r})&=&2\int_0^\infty \frac{d^3p}{(2\pi)^3}f_e({\bf r}; {\bf p}, \mu_e), \label{eq:electron}\\
\rho_{i=p,n}({\bf r})&=&2\int_0^\infty \frac{d^3p}{(2\pi)^3} f_{i}({\bf r}; {\bf p}, \mu_i), \label{eq:baryon}\\
\mu_n&=&\nu_n({\bf r})+g_{\omega N}\omega_0({\bf r})-g_{\rho N}R_0({\bf r}),\label{eq:cpotn}\\
\mu_p&=& \nu_p({\bf r})+g_{\omega N}\omega_0({\bf r})+g_{\rho N}R_0({\bf r})-V_0({\bf r}),\label{eq:cpotp}
\eeqa
We demand that the chemical potentials $\mu_a$ $(a=p,n,e)$ become independent of the position ${\bf r}$.
Note that the condition $\mu_p=\mu_n-\mu_e$ among chemical potentials is applied only in the case of beta-equilibrium.
Here we'd like to make a comment about the inclusion of the Coulomb potential. 
The electron chemical potential as well as the proton chemical potential is the gauge-variant quantity 
and only the charge density is gauge-invariant. 
Hence, the electron density is not uniform in the presence of the Coulomb potential even when the condition $\mu_e=$const. holds everywhere \cite{voskre}.

Accordingly we list in addition some necessary quantities, the scalar density $\rho^{(s)}$ used in Eq.~(\ref{eq:meson1}),
energy density $\varepsilon$ and the entropy density $s$ below:
\beqa
\rho_{i=p,n}^{(s)}({\bf r})&=&\frac{m_N^*(\br)}{\pi^2}H_3(\br; \nu_i, m_N^*(\br))
\\
\varepsilon(\br)&=&\frac{1}{\pi^2}\sum_{i=n,p}\left[H_5(\br; \nu_i,m_N^*(\br))+m_N^{*2}H_3(\br; \nu_i,m_N^*(\br))\right]\nonumber\\
&&+\frac{1}{2}\left(\nabla\omega_0(\br)\right)^2+\omega_0(\br)(\rho_p+\rho_n)-\frac{m_\omega^2}{2g_{\omega N}}\omega_0^2(\br)\nonumber\\
&&+\frac{1}{2}\left(\nabla R_0(\br)\right)^2+g_{\rho N}R_0(\br)(\rho_p(\br)-\rho_n(\br))-\frac{m_\rho^2}{8g_{\rho N}^2}R^2_0(\br)\nonumber\\
&&+\frac{1}{2}\left(\nabla\sigma(\br)\right)^2+\frac{m_\sigma^2}{2g_{\sigma N}^2} \sigma^2(\br)+\frac{bm_N}{3}\sigma^3(\br)+\frac{c}{4}\sigma^4(\br)\nonumber\\
&&+\frac{1}{\pi^2}\int_0^\infty p^3dpf_e(\br; \bp, \mu_e)
+\varepsilon_C(\br)\nonumber\\
s_{a=n,p,e}({\bf r};\mu_a)&=&2\int_0^\infty \frac{d^3p}{(2\pi)^3} \left[f_a({\bf r}; {\bf p}, \mu_a)\log\left( f_a({\bf r}; {\bf p}, \mu_a)\right)\right.\nonumber\\
&&\quad\left.+\left(1-f_a({\bf r}; {\bf p}, \mu_a)\right)\log\left(1-f_a({\bf r}; {\bf p}, \mu_a)\right)\right],
\eeqa
where $\varepsilon_C$ is the contribution from the Coulomb interaction\footnote{
We here discarded the exchange term, which is evaluated in the local-density approximation as 
$\varepsilon_C^{\rm ex}=-3e^2/4(3/\pi)^{1/3}\rho_{\rm ch}(\br)^{4/3}$ \cite{par}.},
\beq
\varepsilon_C(\br)=\frac{e^2}{2}\int_{V}d^3r'\frac{\rho_{\rm ch}(\br)\rho_{\rm ch}(\br')}{|\br-\br'|},
\eeq
and the functions $H_n, (n=3,5)$ are defined by
\beqa
H_n(\br; \nu_i, m)&\equiv& \int_0^\infty\frac{p^{n-1}dp}{(p^2+m^2)^{1/2}}\left(f_i(\br;\bp,\mu)-f_i(\br;\bp,-\mu)\right)\nonumber\\
&\simeq& \int_0^\infty\frac{p^{n-1}dp}{(p^2+m^2)^{1/2}}f_i(\br;\bp,\mu).
\eeqa

To solve the coupled equations of motion (\ref{eq:meson1})-(\ref{eq:cpotp}), we employ the Wigner-Seitz approximation, 
where whole the space is divided into equivalent cells with a geometrical symmetry in one, two or three dimension.
The cell is divided into grid points and the coupled equations of motion are solved numerically.
The size of the WS cell is optimized so that the Helmholtz free energy,  
\beq
F_{WS}=\int_{V_{\rm WS}} d^3r \left(\varepsilon-T\sum_{a=n,p,e}s_a \right)
\eeq
becomes minimum, where $V_{\rm WS}$ denotes the volume of the WS cell.
In the following we consider two cases: one is a given proton fraction case, $Y_p\equiv \int_{V_{WS}}\rho_p(\br)/\int_{V_{WS}}(\rho_p(\br)+\rho_n(\br))$ and the other is $\beta$-equilibrium case with additional condition, $\mu_n=\mu_p+\mu_e$
The value of $\mu_e$ is adjusted at any time step
to maintain the global charge neutrality, $\int_{V_{WS}}d^3r \rho_{\rm ch}(\br)=0$:
we decrease $\mu_e$ when the total charge in a cell is positive and vice versa.
All the above relaxation procedures are carried out simultaneously.
The parameters used in our calculation are fitted to give the saturation properties of
nuclear matter at $T=0$ and properties of finite nuclei in the ground states \cite{maru05},
which are listed in Table I.
\begin{table*}
\caption{
Parameter set used in RMF in our calculation.
With these parameters, the saturation property of nuclear matter is
reproduced: minimum energy per baryon $-16.3$ MeV at $\rho=0.153$ fm$^{-3}$
with the incompressibility $K=240$ MeV.
}
\begin{tabular}{cccccccc}
\hline\\
$g_{\sigma N}$ & 
$g_{\omega N}$ &
$g_{\rho N}$ &
$b$ &
$c$ &
$m_\sigma$ [MeV]&
$m_\omega$ [MeV]&
$m_\rho$ [MeV]\\
\hline\\
6.3935 & 
8.7207 & 
4.2696 & 
0.008659 &
$-0.002421$ &
 400 &
 783 &
 769 
%
\\
\hline\\
\end{tabular}
\end{table*}

\subsection{Bulk Gibbs calculation}
\label{secBulk}

In this section we discuss about bulk properties of nuclear matter without taking into account of electrons
and the finite-size effects. 
We give some isothermal curves in the $P_B$-$Y_p$ plane and isolepthal curves in the $P_B$-$T$ plane by the bulk Gibbs calculation,
where $P_B$ denotes the baryon partial pressure.
Dropping the electric field $V$ and assuming uniform densities $\rho_i, (i=n,p)$ as well as the meson mean-fields, 
the thermodynamic potential density and the energy density render \cite{muller95},
\beqa
\Omega_B/V&\equiv&-P_B\nonumber\\
&=&-\frac{1}{3\pi^2}\sum_{i=n,p}H_5(\nu_i,m_N^*)-\frac{m_\omega^2}{2g_{\omega N}}\omega_0^2-\frac{m_\rho^2}{8g_{\rho N}^2}R^2_0 +\frac{m_\sigma^2}{2g_{\sigma N}^2} \sigma^2+\frac{bm_N}{3}\sigma^3-\frac{c}{4}\sigma^4\nonumber\\
\varepsilon_B&=&\frac{1}{\pi^2}\sum_{i=n,p}\left[H_5(\nu_i,m_N^*)+m_N^{*2}H_3(\nu_i,m_N^*)\right]\nonumber\\
\!\!\!\!&\!\!\!\!+\!\!\!\!&\!\!\!\!\omega_0(\rho_p\!+\!\rho_n)\!-\!\frac{m_\omega^2}{2g_{\omega N}}\omega_0^2\!+\!g_{\rho N}R_0(\rho_p\!-\!\rho_n)\!-\!\frac{m_\rho^2}{8g_{\rho N}^2}R^2_0\!+\!\frac{m_\sigma^2}{2g_{\sigma N}^2} \sigma^2\!+\!\frac{bm_N}{3}\sigma^3\!\!-\!\frac{c}{4}\sigma^4\!,
\label{omegabulk}
\eeqa
with the meson mean-fields,
\beqa
\sigma&=&\frac{1}{m_\sigma^2}\left[g_{\sigma N}(\rho_p^{(s)}+\rho_n^{(s)})-\frac{dU}{d\sigma}\right]\nonumber\\
\omega&=&\frac{g_{\omega N}}{m_\omega^2}(\rho_p+\rho_n)\nonumber\\
\rho&=&\frac{g_{\rho N}}{m_\rho^2}(\rho_p-\rho_n),
\eeqa
where we subtracted a rather trivial electron contribution.
The entropy density is given by Eq.~(\ref{omegabulk}) or follows from the Gibbs-Duhem relation,\beq
s_B\equiv (P_B+\varepsilon_B-\sum_{i=n,p}\mu_i\rho_i)/T,
\eeq 
so that we have the Helmholtz free energy density,
\beq
{\cal F}_B=\varepsilon_B-Ts_B.
\eeq
 

\begin{figure}[h]
\centerline{
\includegraphics[width=.5\textwidth]{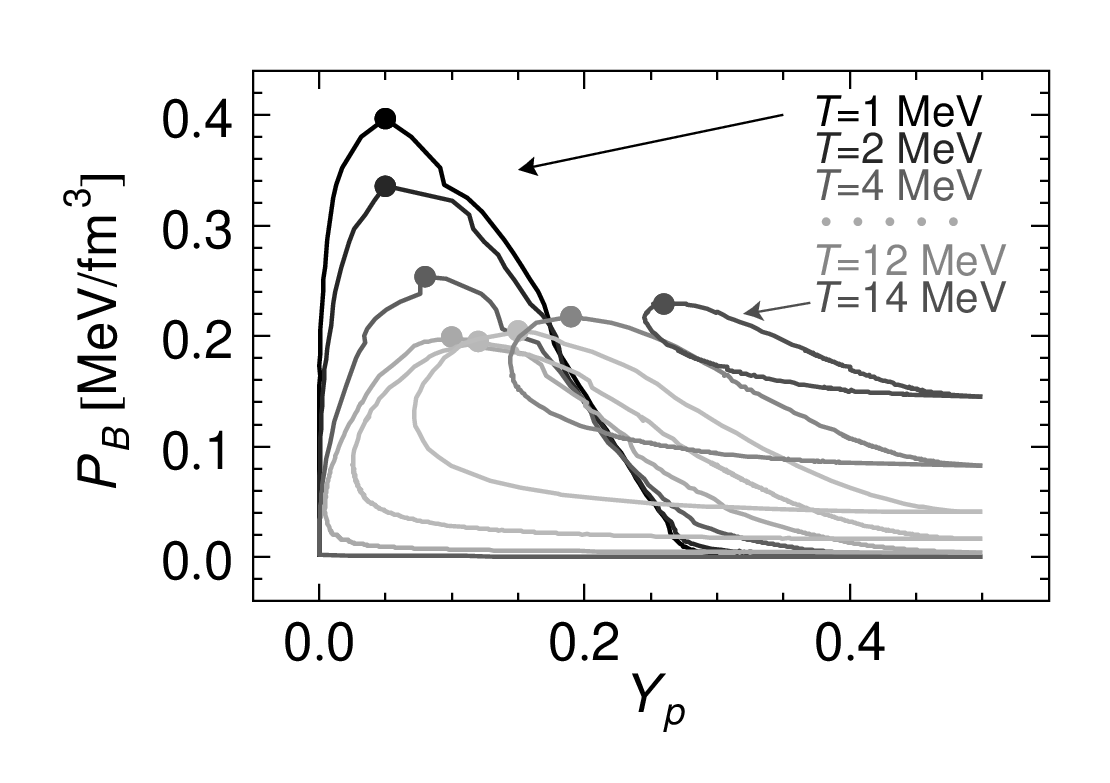}
}
\caption{
Temperature-dependence of phase coexistence curves in $Y_p$-$P_B$ plane for SNM $Y_p=0.5$. 
Filled circles show the critical points.
}
\label{figYPa}
\end{figure}

\begin{figure}[h]
\centerline{
\includegraphics[width=.8\textwidth]{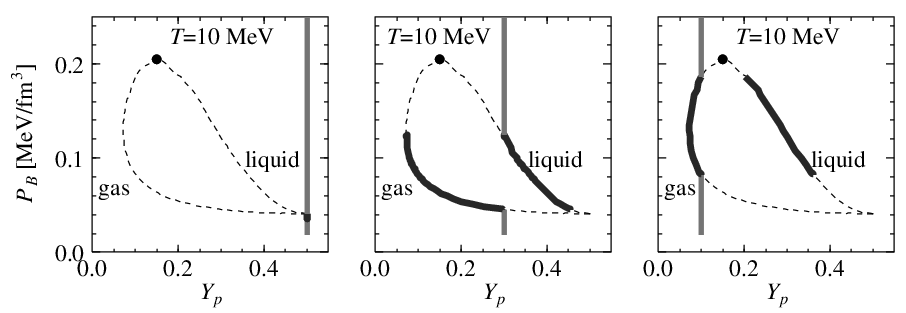}
}
\caption{
Same as Fig.\ \ref{figYPa}  for $T=10$ MeV. 
}
\label{figYPb}
\end{figure}

In Fig.\ \ref{figYPa}, binodal curves with several temperatures are presented in the $Y_p$-$P_B$ plane. 
For a given $Y_p$ and temperature $T$, e.g. $Y_p=0.3$ and $T=10$ MeV (Fig.\ \ref{figYPb},  ASNM is in the gas phase when pressure is small enough. 
As pressure increases the system crosses the binodal curve, where the phase separation occurs and mixed phase begins ($\lambda=0$). 
The gas and liquid phases coexist in the mixed phase and the proton fraction as well as baryon number density in each phase are different. 
As pressure further increases, the volume fraction of each phase $\lambda$ get larger, and eventually total system becomes 
the pure liquid phase ($\lambda=1$). 
The critical point is given by the condition, $dP_B/dY_p|_{\rm CP}=0$, since the two phases cannot be distinguished there, $P_B(T,Y_p^{\rm CP})=P_B(T,Y_p^{\rm CP}+\delta Y_p)$. 
On the other hand, if we dare to apply the Maxwell construction (forced congruent transition), 
we have the congruent curve, where the critical point is the terminal point.

Note that the phase transition is congruent for symmetric nuclear matter ($Y_p=0.5$). 
This is because the free energy should be smallest at $Y_p=0.5$ for every density.
In this case the trajectory of the system and the coexistence curve 
meet at a single point on the $Y_p$-$P_B$ plane
and the mixed phase of this system consists of
gas and liquid with the same value of $Y_p=0.5$. Accordingly $\partial P_B/\partial Y_p=0$ 
at this point (Gibbs-Konovalov theorem \cite{tester}).

In the case of $Y_p=0.1$, the system starts as a gas phase  
 in the panel at the far right of Fig.\ \ref{figYPb}.
With increase of density, the pressure goes up till
$P_B\approx 0.08\ {\rm MeV\cdot fm^{-3}}$.
Then the LG mixed phase appears as in the central panel. 
With further increase of density, the pressure goes up
until $P_B\approx0.18\ {\rm MeV\cdot fm^{-3}}$.
At this point $Y_p$ of the gas phase becomes that of 
the total system, 0.1, which means that the system becomes a gas again. 
However such ``supercritical'' gas has very different properties 
from the usual gas at low pressure.
Such a transition, (gas) $\rightarrow$ (mixed phase) $\rightarrow$ (supercritical gas),
is peculiar to systems with two or more chemical components
and called ``retrograde condensation'' \cite{lan}.

\begin{figure}[h]
\centerline{
\includegraphics[width=.9\textwidth]{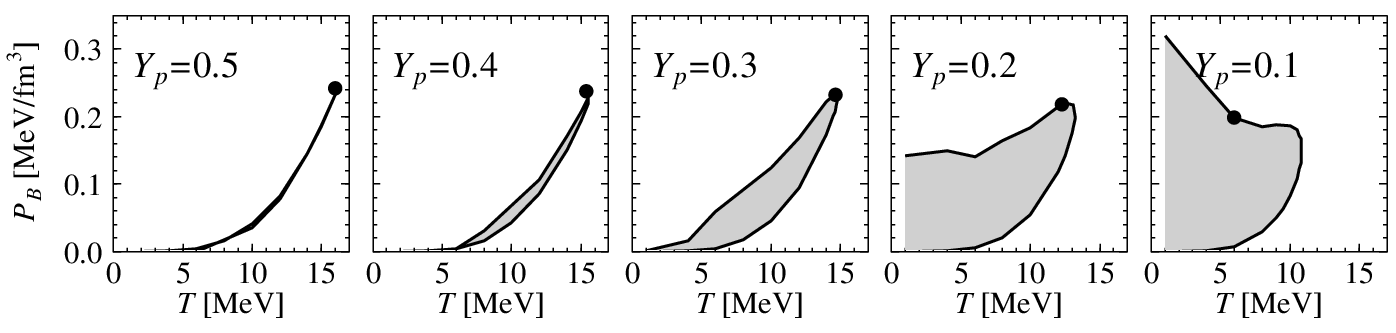}
}
\caption{
Phase coexistence curve on $T$-$P_B$ plane. 
Gray area shows the region of LG mixed phase.
Dots on the curves indicate the critical points.
The panel at the far left exhibits the congruent curve (see Fig. \ref{saturation}), while others the non-congruent ones.}
\label{figTP}
\end{figure}

This situation is more clearly seen in Fig.\ \ref{figTP} , where the binodal curves 
are shown in the $P_B$-$T$ plane for given $Y_p$. 

The critical point ($P_{\rm CP},T_{\rm CP}$) is located on the binodal curve, 
but it is not the special point on the curve. 
The supercritical region ($T>T_{\rm CP}$) also appears in this plane 
for $Y_p=0.1$ and 0.2.  

There is no area inside the coexistence curve for $Y_p=0.5$,  
since the system behaves congruently. 
For smaller $Y_p$, on the other hand, the area become wider.
The gas phase for $Y_p\leq 0.2$ includes dripped neutrons even at $T=0$. 
therefore, the curve is not closed at $T=0$. 




\subsection{Pasta and EOS}

In this section we discuss the non-uniform nuclear matter by taking into account the finite-size effects. 
We include electrons to make the total system to be charge-neutral. 
Electrons are distributed quasi uniformly even in the mixed phase 
and the finite-size effects by electrons is small.
We can see that pasta structures with various geometrical shapes appear in the mixed phase. 
The properties of asymmetric nuclear matter with pasta structures will be elucidated 
by comparing them with those given by the bulk Gibbs calculation in the previous section.


\begin{figure}[h]
\centerline{
\includegraphics[width=.28\textwidth]{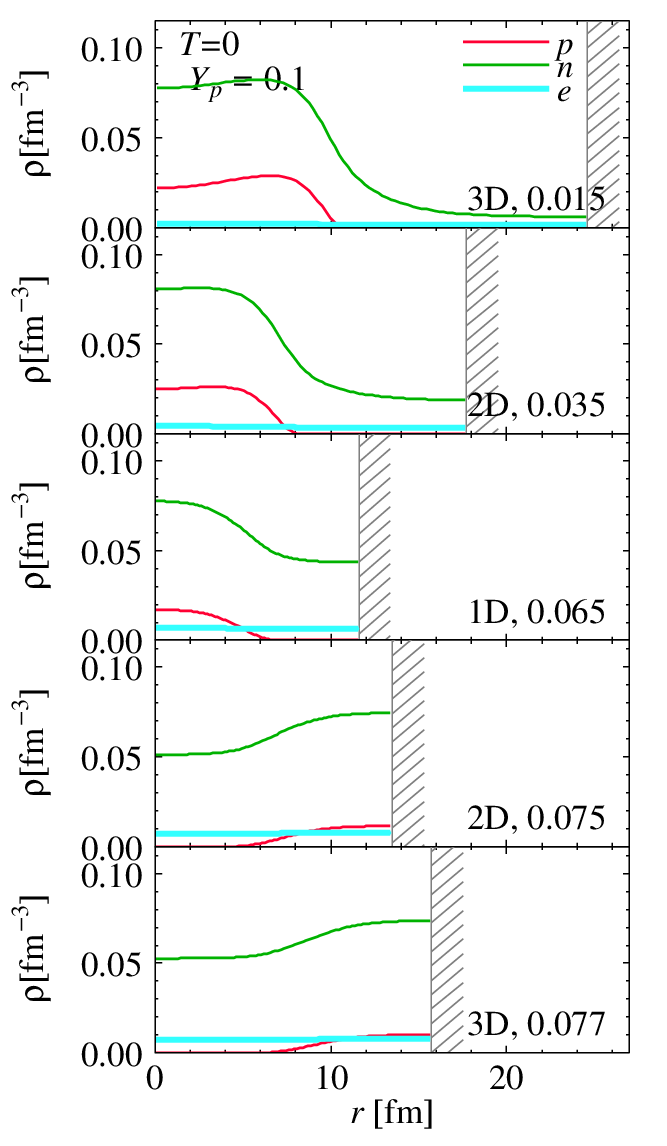}
\includegraphics[width=.28\textwidth]{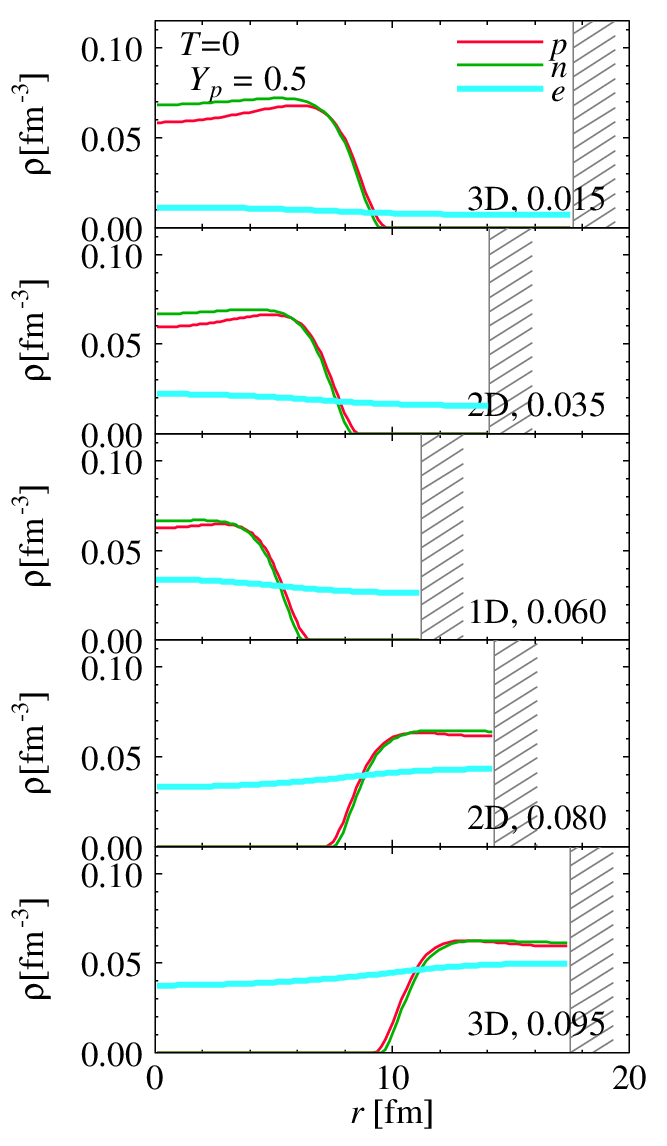}
\includegraphics[width=.28\textwidth]{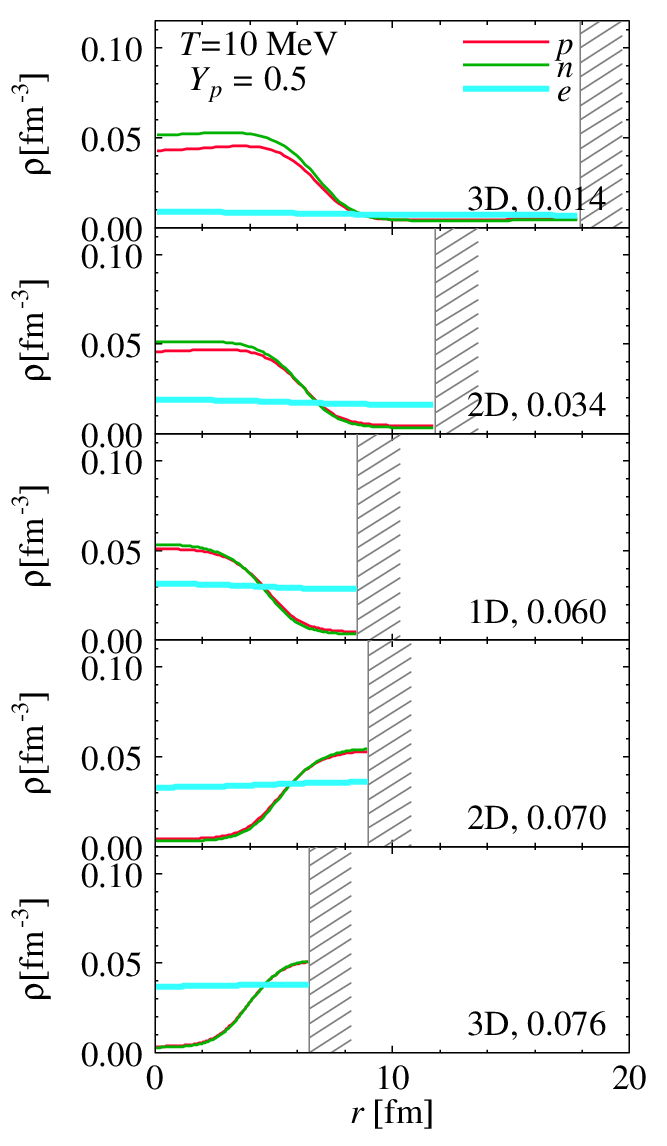}
}
\caption{
Examples of the density profiles in the Wigner-Seitz cells.
with $Y_p=0.1$ (left), $Y_p=0.5$ $T=0$ (middle) and  $T=10$ MeV (right).
}
\label{figPasta}
\end{figure}


We show in Fig.\ \ref{figPasta} some typical density profiles in the cell. 
The left and the middle panels show the cases of 
proton fraction $Y_p=0.1$ and $Y_p=0.5$.
Apparently, dense nuclear phase (liquid) and dilute nuclear/electron 
phase (gas) are separated in space and they form pasta structures 
depending on density.
One should notice that coexisting two phases 
are non-congruent and have different components, 
i.e.\ nuclear matter and electron gas.
Therefore the EOS of the whole system cannot be obtained
by the Maxwell construction.
Since electron density is almost uniform 
and independent of baryon density distribution,
we can separately discuss the properties of the baryon partial system.

\begin{figure}[h]
\centerline{
\includegraphics[width=.42\textwidth]{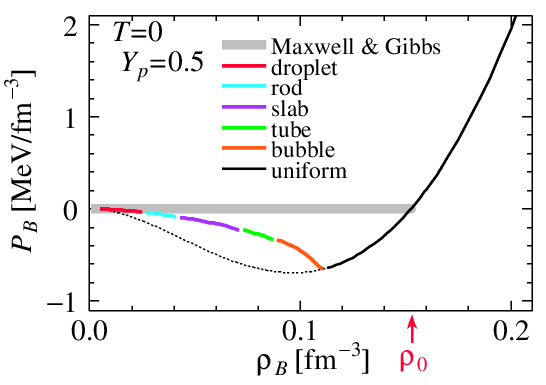}
\includegraphics[width=.42\textwidth]{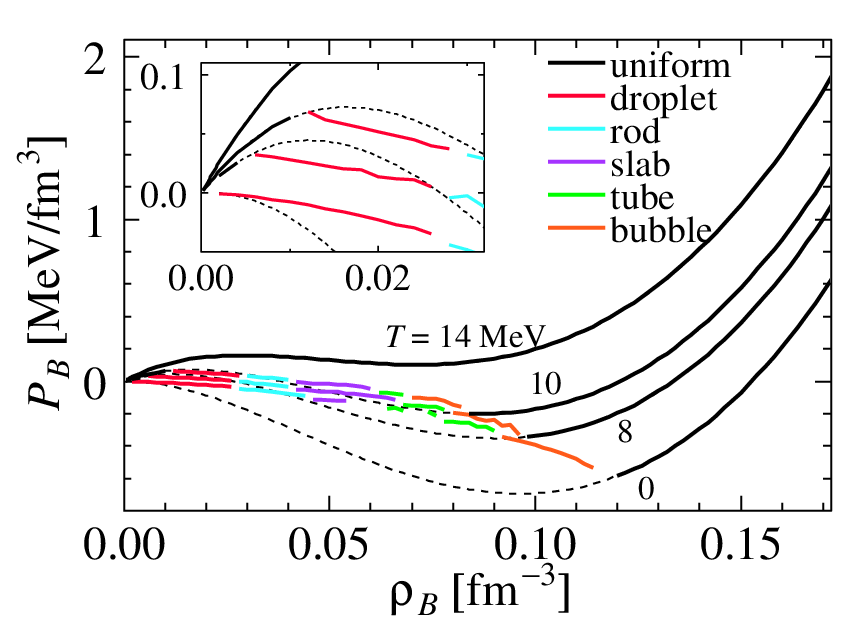}
}
\caption{
Baryon partial pressure as a function of density $\rho_B$
for symmetric nuclear matter $Y_p=0.5$, 
which is given by subtracting electrons pressure from the total one.
\vspace{-5mm}}
\label{figEOS05}
\end{figure}

In the case of $Y_p=0.5$, as discussed in Sec.\ \ref{secBulk},
the local proton fraction stays almost constant.
Therefore the system behaves like a system with single component.
This means that one can apply the Maxwell construction to get 
the baryon partial pressure $P_B$ as in the left panel 
of Fig.\ \ref{figEOS05}:
uniform low-density matter with a negative partial pressure is 
not favored and the Maxwell construction gives $P_B=0$ for the mixed phase. 
By the finite-size effects, i.e.\ the Coulomb repulsion and 
the surface tension, the structured mixed phase becomes unstable 
in the density region just below $\rho_0$,
and consequently uniform matter with a negative partial pressure is allowed. 
Note again that the total pressure including electrons is always positive 
even in this case, so that the system is thermodynamically stable.

In the case of asymmetric nuclear matter, 
e.g.\ $Y_p=0.1$ in the left panel of Fig.\ \ref{figPasta},
the proton fraction in the dilute and dense phases are 
different, especially for low $Y_p$.
Matter behaves as a system with multi chemical components
and the Maxwell construction does not satisfy the Gibbs conditions.





\subsection{Thermal effects}

Let us discuss the thermal effects on the LG mixed phase of 
low density nuclear matter.
By comparing the density profiles in the middle and 
right panels of Fig.\ \ref{figPasta}, we easily notice that 
the dilute phase at finite temperature always contains baryons 
while they are absent in the dilute phase at zero temperature 
if $Y_p\approx 0.5$.
This is due to the Fermi-Dirac distribution at finite temperature, 
where density as a function of chemical potential is always positive.

We also notice that the size of the pasta structure is smaller
in the case of finite temperature.
This comes from a reduction of the surface tension between two phases
at finite temperature since the difference of baryon densities  
between two phases get smaller.

The isothermal EOS's (baryon partial pressure as a function of 
baryon number density) of symmetric nuclear matter 
at various temperatures are shown in the left panel of Fig.\ \ref{figEOS05}.
Dotted and thick solid curves show the cases of uniform matter,
while thin solid curves are the cases where non-uniform pasta structures 
are present.
As shown in the right panel of Fig.\ \ref{figEOS05},
pasta structures appear at finite temperatures as well as the case of $T=0$.
But there appears uniform matter (gas phase) at the lowest-density region
\cite{avancini,friedman}
since the baryon partial pressure of uniform matter has
a positive gradient against density.
On the other hand, the uniform matter is unstable
where the pressure gradient is negative
even if the pressure itself is positive.
At $T=14$ MeV, we obtain no pasta structure
since the baryon partial pressure of uniform matter
becomes a monotonic function of density above this temperature.
Study of the instability of uniform matter and the appearance 
of the pasta structures  in connection with 
the spinodal region is in progress. 

In Sec.\ \ref{secBulk} we have shown a retrograde condensation
at $T=10$ MeV and $Y_p=0.1$ by the bulk calculation.
In the full calculation with pasta structures, however, 
there is no evidence to have such phenomenon so far.
Probably it is washed away by the finite-size effects.



\subsection{Discontinuity in pressure}

If one applies bulk Gibbs conditions to the low-density nuclear matter,
the pressure changes continuously with the density.
Mechanically unstable region with negative gradient of pressure $dP/d\rho_B<0$ 
never appears.
However, with inclusion of the finite-size effects for inhomogeneous nuclear pasta,
some part of density region becomes unstable for the mixed phase.
In this density region, uniform matter with negative baryon pressure is allowed
(see Fig.\ \ref{figEOS05}).
The baryon pressure of the mixed phase also can be negative.
However, the total pressure including that of electron 
and its gradient are always positive.

Another thing to note is that there are discontinuities in 
pressure at the densities where the geometrical structure changes.
The free-energy density as a function of density is continuous 
since it is the criterion of choosing the optimal structure.
On the other hand the pressure $P_B=\sum_i\mu_i\rho_i-\varepsilon$
is not necessarily continuous since the chemical potential $\mu_i$
should be dependent on the structure.
So far we employ the Wigner-Seitz cell approximation for matter with the pasta structures.
This is rather a strong constraint on the structure.
In the work of Ravenhall et al.\ \cite{Rav83} dimension parameter
is optimized to get the ground state structure. 
It behaves smoothly at the transient region
except for a jump between uniform and bubble.
If we could consider more than the typical pasta structures \cite{oka},
the discontinuity of pressure may diminish.

\section{Hadron-quark deconfinement transition}\label{secHQ}

In this section we discuss the emergence of the pasta structure associated 
with the hadron-quark deconfinement transition\footnote{
We, hereafter, dismiss the possibility of color superconductivity \cite{ARRW}.}.

Nowadays the hadron-quark deconfinement transition and its implications 
has been extensively studied by various viewpoints: 
a great deal of effort has been paid for quest of signals in relativistic heavy-ion collisions \cite{QCD}, 
compact-star phenomena \cite{bec} or remnants of 
primordial QCD transition \cite{uni}. Theoretically, numerical simulations based on the lattice gauge theory 
have been done to estimate the critical temperature and figure out the properties of the phase transition.
However,  
for the present, any numerical method is useless for the deconfinement in compact stars. Instead, the 
studies by using the effective models of QCD 
such as the MIT bag model or the NJL model have been done. 

Although the order of the phase transition is not clear, 
theoretical arguments have suggested it should be of the first order in the low temperature and high-density regime \cite{QCD}. 
So, let us assume here it is of the first order. 
Then we can expect the emergence of the structured mixed phase as in liquid-gas transition. 
Actually Glendenning first discussed it to demonstrate 
how the Maxwell construction is insufficient in neutron-star matter. Before the appearance of his paper,  
when the first-order phase transition occurs and new phase appears, 
EOS has been constructed by simply applying the Maxwell construction. 
Using the bulk calculation without the Coulomb interaction or the surface energy, called the finite-size effects,  
he could show a wide density region of the mixed phase for hadron-quark deconfinement transition, 
the pressure of which is by no means constant.   

Once the finite-size effects are properly taken into account, inhomogeneous matter called pasta emerges as is already shown in the previous section.
Heiselberg et al.\ immediately pointed out the importance of the finite-size effects and demonstrated 
how the surface energy parametrized by the surface-tension parameter makes the region of the structured mixed phase narrower \cite{HPS93}: 
they showed the disappearance of the structured mixed phase for a critical value of the surface tension. 
They considered only droplets for hadron-quark deconfinement transition, 
while more general pasta structures or the crystalline structures have been subsequently discussed \cite{glenpei}.
Voskresensky et al.\ discussed the charge screening effect on the stability of the structured mixed phase during the deconfinement transition, 
and showed its relevance for a given surface tension \cite{voskre}. 
They emphasized that charge screening and rearrangement of particle densities become important for the description of the mixed phase and EOS. 
Using the effective model for hadron matter and the MIT bag model for quark matter, 
Endo et al.\  explicitly demonstrated these features \cite{end05}.  

The surface tension should be calculated self-consistently with models of hadron and quark matter. 
Unfortunately, there is a large ambiguity about its value, whereas a large value of $100$MeV$\cdot$fm$^{-2}$ might be more realistic \cite{ARRW}.

\subsection{Framework using Brueckner-Hartree-Fock calculation and the MIT bag model}
There are many models to describe the EOS with hyperons.
However, nowadays, the potentials of the baryon-baryon interactions have been directly calculated by the great successes in the lattice QCD calculations~\cite{nemura09}. 
Hence, it is very hot topic to calculate the EOS from the baryon-baryon interaction directly.
Here, we show the EOS by the theoretical framework of the nonrelativistic Brueckner-Hartree-Fock approach~\cite{baldo98a} based on the microscopic nucleon-nucleon~($NN$), nucleon-hyperon~($NY$). The Brueckner-Hartree-Fock calculation is a reliable and well-controlled theoretical approach for the study of dense baryonic matter.
Detailed procedure can be found in Refs.\ \cite{baldo98a,baldo98b,schulze95,vidana95}.

We adopt the Argonne $V_{18}$ potential~\cite{wiringa95} for $NN$ potentials, 
and semi-phenomenological Urbana UIX nucleonic three-body forces~\cite{pudliner97}     
and the Nijmegen soft-core NSC89 $NY$ potentials \cite{maessen89}.
Unfortunately, there are not reliable potentials for the hyperon-hyperon interaction now. 
Therefore, we neglect them here.
We strongly hope that the lattice QCD calculations or the experiments, such as J-PARC, will reveal the these interactions in the future.

Moreover, we discuss the hadron-quark mixed phase with the finite-size effects at finite temperature
with/without trapped neutrinos. To take into account the effects of finite temperature, we adopt \textit{Frozen Correlations Approximation}~\cite{baldo99, nicotra06a,nicotra06b}.
In this approximation, the correlations at finite temperature are assumed to be the same with the ones at zero temperature. 
It is found to be good accuracy at finite temperature
by past studies~\cite{baldo99, nicotra06a,nicotra06b}. 
Accordingly, we focus on the hadron-quark mixed phase at finite temperature 
and adopt this approximation.

In the following, we briefly describe our framework.
First we get the chemical potential $\mu_i$  from the number density $n_i$, 
\begin{eqnarray}
\rho_i     &=& \cfrac{g}{(2\pi)^3} \int^\infty_0 f_i(p) ~ 4 \pi p^2 dp     \\
f_i(p) &=& \cfrac{1}{ {\rm exp} \{ (\varepsilon_i-\mu_i)/ T \} +1},
\end{eqnarray}
where  $\varepsilon_i$ and $f_i(p)$ are the single-particle energy and the Fermi-Dirac distribution function, respectively, whereas subscript $i$ denotes the particle species, $i=n,p,\Lambda,\Sigma^-$. 
We set each degeneracy factor $g=2$, and adopt each mass as $m_n=m_p=939$ MeV, $m_\Lambda=1115.7$ MeV, and $m_{\Sigma^-}=1197.4$ MeV.
We note that $\varepsilon_i$ includes the interaction energy $U_i$ as well as the kinetic energy \cite{baldo99, baldo04}, 
\begin{eqnarray}
\varepsilon_i = \sqrt{m_i^2 + p^2 } + U_i.
\end{eqnarray}

Finally, we get the free-energy density $\mathcal{F}$ as following   
\begin{eqnarray}
\mathcal{F}_H = \sum_i  \left\{ \cfrac{g}{(2\pi)^3} \int^\infty_0 \sqrt{m_i^2+p^2} f_i(p) ~4 \pi p^2 dp + \cfrac{1}{2} ~ U_i \rho_i \right\}-Ts_H ,
\label{eq:free}
\end{eqnarray}
where $s_H$ is the entropy density calculated from
\begin{eqnarray}
s_H = - \sum_i \cfrac{g}{(2\pi)^3} 
\int^\infty_0 \{ f_i(p) {\rm ln}f_i(p) +(1-f_i(p)) {\rm ln}(1-f_i(p)) \} ~ 4 \pi p^2 dp.
\end{eqnarray}

The total pressure for uniform hadron phase is given by
\begin{eqnarray}
P_H= \sum_i \mu_i \rho_i   -  \mathcal{F}_H. 
\end{eqnarray}

Next, we show the quark matter based on the thermodynamic bag model.
In this model, we get the number density $n_Q$, the pressure $P_Q$, and the energy density $\epsilon_Q$,
\begin{eqnarray}
\rho_Q&=&\sum_q \cfrac{g}{(2\pi)^3} \int^\infty_0 f_q(p) 4 \pi p^2 dp, \\ 
\epsilon_Q&=&\sum_q \cfrac{g}{(2\pi)^3} \int^\infty_0 \sqrt{m_q^2 + p^2 } f_q(p) 4 \pi p^2 dp +B, \\ 
s_Q &=& - \sum_q \cfrac{g}{(2\pi)^3} 
\int^\infty_0 \{ f_q(p) {\rm ln}f_q(p) +(1-f_q(p)) {\rm ln}(1-f_q(p)) \} ~ 4 \pi p^2 dp \\
\mathcal{F}_Q&=& \epsilon_Q -Ts_Q \\
P_Q&=& \sum_q \mu_q \rho_q   -  \mathcal{F}_Q. 
\end{eqnarray}
for uniform quark matter,
where $f_q(p)$ is the Fermi-Dirac distribution function of the quark $q$ ($=u,d,s$), $m_q$ its current mass, 
and $B$ the energy-density difference between 
the perturbative vacuum and the true vacuum, the bag constant $B$ set to be 100~MeV$\cdot$fm$^{-3}$.

However, there are many uncertainties for $B$. Here, we also adopt the another type of $B$ by the density dependent bag model. In this model, we set the vacuum energy $B(\rho_B)$ as 
\begin{eqnarray}
 B(\rho_B) = B_\infty +(B_0 - B_\infty) \exp[-\beta \frac{\rho_B}{\rho_0}] \nonumber
 \label{eq:01}
 \end{eqnarray}   
with $B_\infty = 50$ MeV$\cdot$fm$^{-3}$, $B_0=400$ MeV$\cdot$fm$^{-3}$, and $\beta=0.17$. Here, we set the saturation density as $\rho_0=0.17$ fm$^{-3}$. This approach has been proposed in Ref.\ \cite{nicotra06b}, on the basis of experimental results obtained at CERN SPS on the formation of a quark-gluon plasma. The just mentioned choice of the parameters allows symmetric nuclear matter to be in the pure hadronic phase at low densities, and in the quark phase at large densities, while the transition density is taken as a parameter.

For the mixed phase, we must numerically search the minimum of the free energy
\begin{eqnarray}
F = \int_{V_H} dr^3 \mathcal{F}_H[n_i] 
 + \int_{V_Q} dr^3 \mathcal{F}_Q[n_q] + F_e + F_{\nu_e}+ E_C + E_S
\label{eq:03}
\end{eqnarray}
changing the volume ratio of quark matter to hadron matter, the geometrical types of structures, 
and the size of Wigner-Seitz cell $R_W$. Here, the free energies of electrons, $F_e$, and neutrinos, $F_{\nu_e}$, are calculated by the use of the Fermi-Dirac distribution functions. We do not consider all kinds of anti-particles for simplicity.
 
In the above equation, $E_S$ stands for the surface energy which comes from
a sharp boundary between hadron and quark phases with a fixed surface tension $\tsurf$.
We demand the global charge neutrality and the chemical equilibrium as,
\begin{eqnarray} \label{eq:02}
 && \mu_u+\mu_e - \mu_{\nu{_e}}= \mu_d = \mu_s , \nonumber\\
 && \mu_p+\mu_e - \mu_{\nu{_e}}= \mu_n = \mu_\Lambda = \mu_u + 2\mu_d ,\\
 && \mu_{\Sigma^-} + \mu_p = 2\mu_n .\nonumber
\end{eqnarray}

Although the surface tension of the hadron-quark interface is poorly known, 
some theoretical estimates based on the MIT bag model 
for strangelets \cite{jaf} and
lattice gauge simulations at finite temperature \cite{latt} suggest
a range of $\tsurf \approx 10$--$100\;\rm MeV\cdot fm^{-2}$.
We will discuss the effects of surface tension on the pasta structures in subsection 1.4.4.

\begin{figure}
\begin{center}
\includegraphics[width=0.48\textwidth]{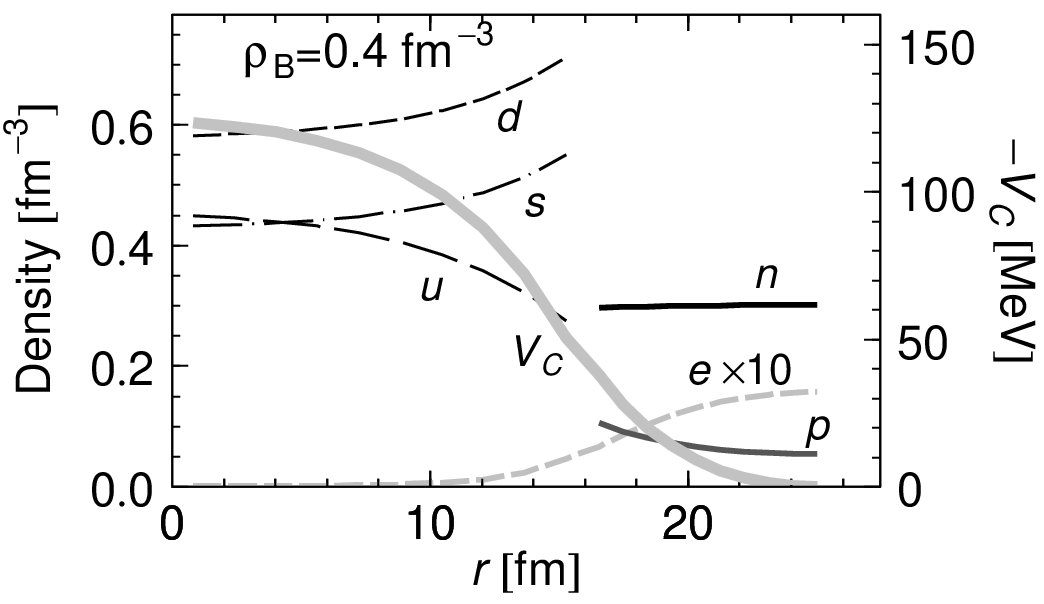}
\end{center}
\caption{
Density profiles 
and Coulomb potential $V_C$ 
within a 3D (quark droplet) Wigner-Seitz cell
of the mixed phase at $\rho_B=0.4$ fm$^{-3}$.
The cell radius and the droplet radius are $R_W=26.7$ fm
and $R=17.3$ fm, respectively.
}
\label{figProf}
\end{figure}

Fig.~\ref{figProf} shows an example of the resulting density profile in the droplet phase for $\rho_B=0.4$ fm$^{-3}$ at $T=0$.
One can see the non-uniform density distributions of particle species 
together with the non-vanishing Coulomb potential.
The quark phase is negatively charged, so that 
$d$ and $s$ quarks are repelled to the phase boundary, 
while $u$ quarks gather at the center.
The protons in the hadron phase are attracted by the negatively charged 
quark phase, while the electrons are repelled.

\subsection{Hadron-quark mixed phase and compact stars}
\begin{figure}[h]
\vspace{-3mm}
\begin{center}
\includegraphics[width=.5\textwidth]{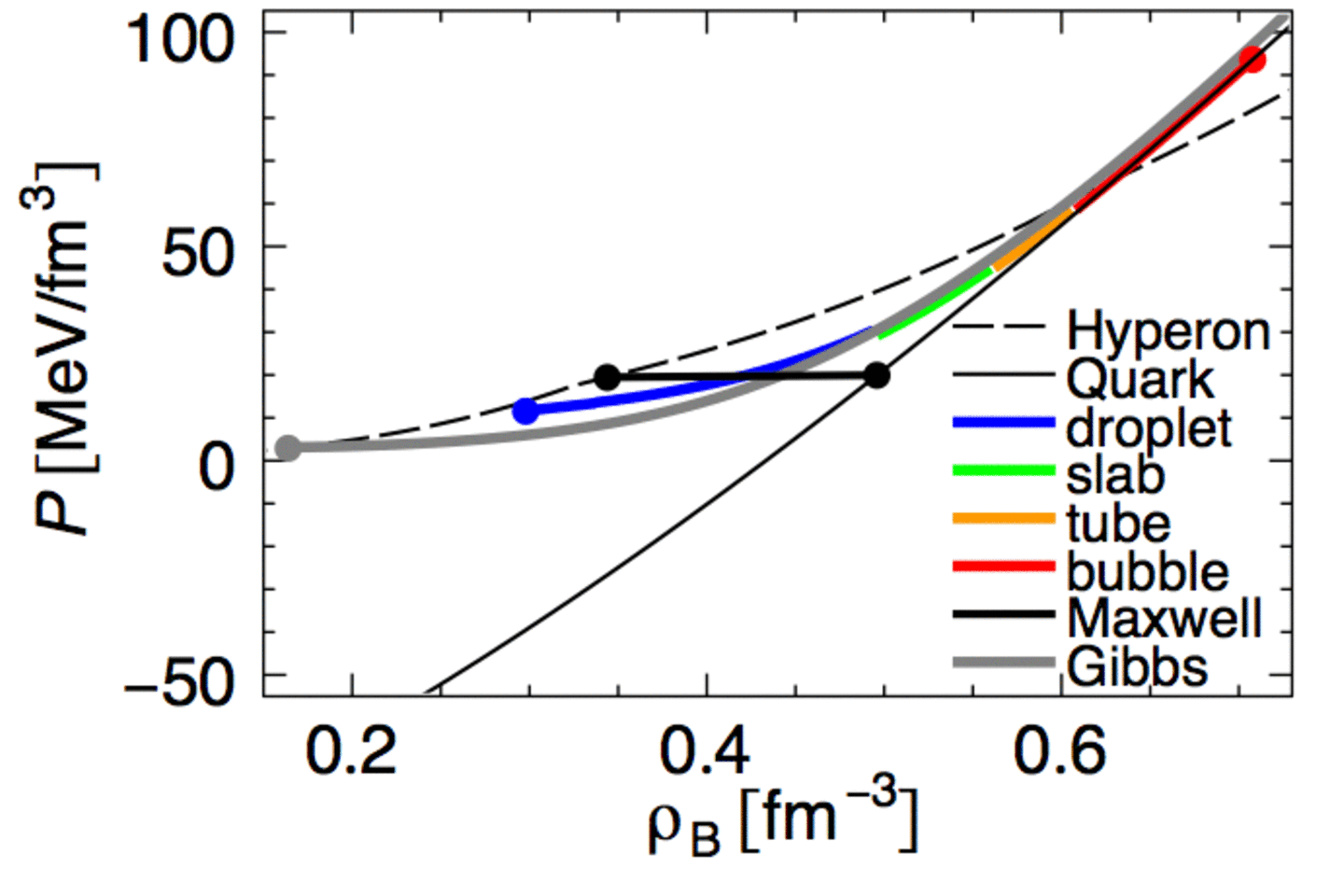}
\includegraphics[width=0.48 \textwidth]{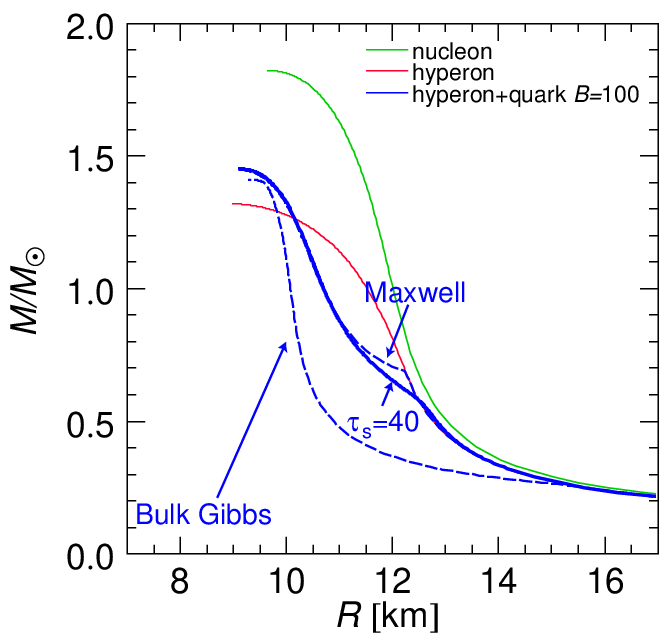}
\end{center}
\caption{
Left panel shows the EOS of the mixed phase (thick dots) in comparison with pure hadron and quark phases (thin curves) at zero temperature. We also show, for comparison, the mixed phase by the Maxwell construction by thin solid line. Note that the Maxwell construction is incorrect in the present case, but it is still useful as an eye guideline.
Right panel show mass-radius relation of calculated neutron stars.
}
\label{figMRrelation}
\end{figure}

Once obtained the EOS of matter, we can know the density profile 
(density as a function of radial distance from the center) and consequently
the mass and the radius of compact stars.
This is done by numerically solving the Tolman-Oppenheimer-Volkoff (TOV) 
equations \cite{sha} as
\begin{eqnarray}
  {dp\over dr} &=& -{ G m \epsilon \over r^2 } \,
  {  \left( 1 + {p / \epsilon} \right) 
  \left( 1 + {4\pi r^3 p / m} \right) 
  \over
  1 - {2G m/ r} } \:,\qquad
\label{tov1}\\
  {dm \over dr} &=& 4 \pi r^2 \epsilon \:,
\label{tov2}
\end{eqnarray}
being $G$ the gravitational constant. 
Starting with a central mass density $\epsilon(r=0) \equiv \epsilon_c$,  
one integrates out Eqs.~(\ref{tov1}) and (\ref{tov2}) until 
the surface pressure becomes to zero.
This gives the stellar radius $R$ and its gravitational mass $M=m(R)$.

We show the relation between EOSs and structures of compact stars 
in Fig. \ref{figMRrelation}. The left panel is the EOSs and the right panel 
 shows the mass-radius relation of calculated neutron stars.
As for the EOSs, we shows three types of quark-hadron phase transitions;
 c.g. the EOSs by the bulk Gibbs condition, the Maxwell construction, and 
full calculation considering the finite size effects.
Here, the surface tension is set as $\tsurf=40$ MeV$\cdot$fm$^{-2}$. 
We also show the uniform hadron and quark matters for comparison.
Using these EOSs and the TOV equations, 
we can calculate the mass-radius relations shown in the right panel. 
It is well-known that the observed masses of neutron stars are distributed
 in a narrow band close to 1.4 times the solar mass $M_\odot$ \cite{nsmass}. 
The maximum mass of neutron star is enough larger than the observed masses.
With inclusion of hyperon degree of freedom, however, 
the softening of matter reduces the possible mass of neutron stars 
as low as 1.3 $M_\odot$, which contradicts the observation.
The blue curves show the cases with hadron-quark mixed phase.
The upper most one and the bottom are the results with the EOSs obtained 
by the Maxwell construction and the bulk Gibbs calculation, respectively.
The result with inclusion of the full pasta structures 
with $\tsurf=40$ $\rm MeV \cdot fm^{-2}$ lies just below the Maxwell construction case. 

In any case with the hadron-quark mixed phase, the calculated maximum mass 
is slightly above $1.4 M_\odot$, which might be tolerable in confrontation with the observation.
What is important is that the mixture of hyperons in nuclear matter
softens the EOS, which causes the contradiction with the observation, 
i.e.\ the ``maximum-mass problem''.
By considering the deconfinement transition to quark matter, the EOS gets stiffer at
higher density region. 
This stiffness at high density sustains the neutron stars with $M\sim 1.4 M_\odot$.

\subsection{Hyperon suppression in the mixed phase}

Though we allow the freedom of hyperon mixture, hyperons don't actually 
come into the mixed phase.
\begin{figure}[t]
\begin{center}
\includegraphics[width=0.7\textwidth]{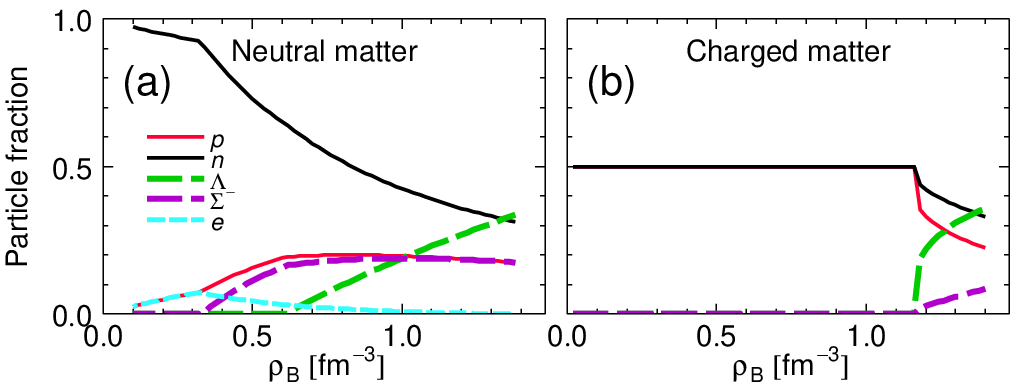}
\end{center}
\vspace{-3mm}
\caption{
(a) Particle fractions of neutral matter with electrons. 
(b) The same quantity for charged matter without electrons,
the low-density part of which corresponds to symmetric nuclear matter.
}
\label{figRatioUnif}
\end{figure}
The suppression of hyperon mixture in the mixed phase
is due to the fact that the hadron phase is positively charged.
As shown in Fig.\ \ref{figRatioUnif}, 
hyperons ($\Sigma^-$) appear in charge-neutral hadronic matter 
at low density 
to reduce the Fermi energies of electron and neutron.
Without the charge-neutrality condition, on the other hand, 
there is symmetric nuclear matter at lower density
and hyperons will be mixed there 
at higher density due to the large hyperon masses.
Generally speaking hyperons are hardly mixed in the positively charged matter.
Thus, the mixture of hyperons is suppressed in the mixed phase where the hadron phase 
is positively charged.

\subsection{Effects of surface tension}

\begin{figure}[h]
\vspace{-3mm}
\begin{center}
\includegraphics[width=0.48\textwidth]{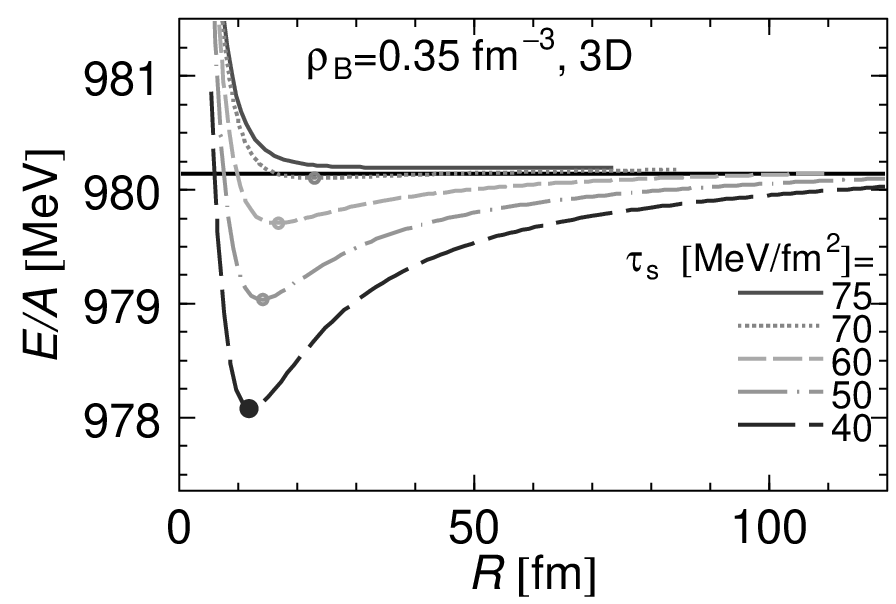}
\end{center}
\caption{
Droplet radius ($R$) dependence of the energy per baryon
for fixed baryon density $\rho_B=0.35$ fm$^{-3}$
and different surface tensions.
The temperature $T$ is zero for all cases.
The quark volume fraction $(R/R_W)^3$ is fixed for each curve.
Dots on the curves show the local energy minima.
The black line shows the energy of the Maxwell construction case.
}
\label{figRdep}
\end{figure}

Let us consider the role of the surface tension on the mixed phase.
As already mentioned, the surface tension between hadron and quark phases is unknown.
If one uses a smaller surface tension parameter $\tsurf$, 
the energy gets lower and the density range of the mixed phase gets wider.
The limit of $\tsurf=0$ leads to a bulk application of 
the Gibbs conditions without the Coulomb and surface effects.  
On the other hand, using a larger value of $\tsurf$, 
the geometrical structures increase in size and
the EOS gets closer to that of the Maxwell construction case. 
Above a limiting value of $\tsurf \approx 65\;\rm MeV\cdot fm^{-2}$ 
the structure of the mixed phase becomes mechanically unstable\cite{voskre}: 
for a fixed  volume fraction $(R/R_W)^3$
the optimal values of $R$ and $R_W$ diverge
and local charge neutrality is recovered in the mixed phase, 
where the energy density equals that of the Maxwell construction 
(see Fig.\ \ref{figRdep}).

\subsection{Thermal and neutrino trapping effects}
\begin{figure}[h]
\centerline{
\includegraphics[width=.5\textwidth]{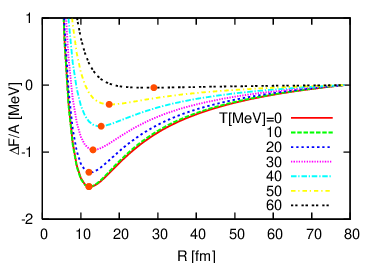}
\includegraphics[width=.5\textwidth]{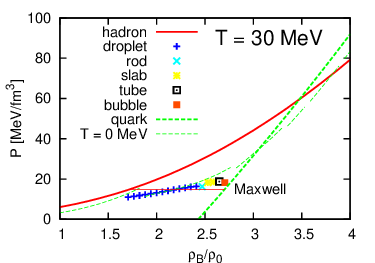}
}
\caption{Left panel shows that the radius $R$ dependence of the free-energy per baryon for the droplet case at $\rho_B = 2~\rho_0$ and different temperatures. The free energy is normalized by the value at $R \rightarrow \infty$. The filled circles on each curve shows the energy minimum. The results are for $B = 100$ MeV$\cdot$fm$^{-3}$, $\tsurf= 40$ MeV$\cdot$fm$^{-2}$. 
The right panel shows the EOS of the mixed phase (thick dots) in comparison with pure hadron and quark phases (thin curves) at $T = 30$ MeV. We also show, for comparison, the mixed phase by the Maxwell construction by thin solid line.}
\label{fig:ny02}
\end{figure}

In following we discuss the thermal and neutrino trapping effects on the quark-hadron phase transition. In the left panel of Fig.\ \ref{fig:ny02} we show the dependencies of the isothermal free-energy on the cell size for several temperatures in Eq.~(\ref{eq:02}) at  $\rho_B = 2~\rho_0$, using the constant bag constant and the surface tension as $B=100$ MeV$\cdot$fm$^{-3}$ and $\tsurf= 40$ MeV$\cdot$fm$^{-2}$, respectively. 

Here we assume the non-uniform structure is the droplet and do not take into account neutrinos to see the thermal effects. The quark volume fraction $(R/R_W)^3$ is fixed to be the optimal value at $T = 0$ MeV for each curve. From the this panel, we can see that the minimum point is shifted to a larger value of $R$ as temperature is increased, and eventually disappears for $T > 60$ MeV.
The temperature dependence of the free energy comes from the Coulomb energy, the surface energy and the correlation energy. By comparing the temperature dependence of these contributions, we can see that the correlation energy is primarily responsible to the behavior of the minimum point. It means that the mixed phase becomes less stable as  temperature is increased ~\cite{yas09}.

\begin{figure}[h]
\centerline{
\includegraphics[width=.5\textwidth]{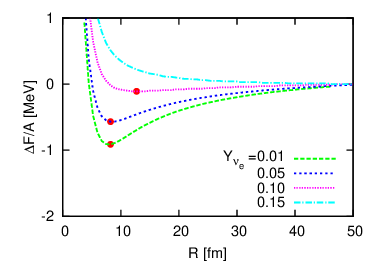}
\includegraphics[width=.5\textwidth]{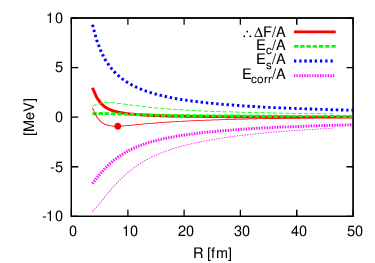}
}
\caption{\label{fig:ny03} Left panel is the same as the left panel of Fig.\ \ref{fig:ny02}, but for different neutrino fractions $Y_{\nu_e}$ at a given temperature, $T=10$ MeV.  In this case, we set baryon density as $\rho_B = 2.5~\rho_0$. The thick lines of the right panel are the case of $Y_{\nu_e} = 0.15$, the thin lines $Y_{\nu_e} = 0.01$.
The right panel shows each component of the free energy ($\Delta F/A$); e.g. the free energy, the Coulomb energy~($E_{\rm C}$/A), the surface energy~($E_{\rm S}$/A), and the correlation energy~($E_{\rm corr}$/A). Here they are shown per baryon.  }
\end{figure}

After the minimization of the free energy, we can get the thermodynamic quantities, such as pressure, entropy, etc..@In the right panel of Fig. \ref{fig:ny02}, we show the EOS (baryon density vs. pressure for the neutrino-free case at $T=30$ MeV. Clearly, the pressure becomes to close to the one given by the Maxwell construction at finite temperature, since the mixed phase becomes unstable and its density regime is thereby largely restricted.
In Fig.\ \ref{fig:ny03}, we show the dependencies of neutrino trappings on the free energy of the mixed phase. Here, we adopt the density-dependent bag constant, and fixed the quark volume fraction $(R/R_W)^3$ at $Y_{\nu_e} = 0.01$ for each curve. For $Y_{\nu_e} > 0.1$, the minimum points disappear. The right panel shows the each component of the free energy.  From this figure, we can see that both the correlation energy and the Coulomb energy mainly contribute to the behavior of the minimum point in the presence of neutrinos, while it is a only  the correlation energy for the temperature dependence.

\begin{figure}[h]
\centerline{
\includegraphics[width=.5\textwidth]{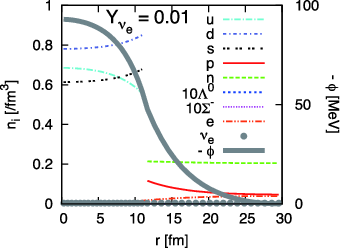}
\includegraphics[width=.5\textwidth]{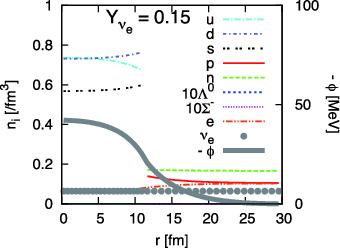}
}
\caption{\label{fig:ny04}Density profiles and Coulomb potential $\phi$ within 1D (slab) for $\rho_B = 2.5~\rho_0$ at $T=10$ MeV. Here, the neutrino fractions are set to  $Y_{\nu_e}=0.01$~(left panel) and $Y_{\nu_e}=0.15$~(right panel). The cell sizes are $R_W = 30$ fm in these figures. The slab size are $R = 10.9 $ fm. $(R/R_W)^3$ is fixed to be the optimal value at $Y_{\nu_e} = 0.01$.}
\end{figure}

To elucidate the neutrino-trapping effect in the mixed phase, we show the density profiles and the Coulomb potential for the slab case in Fig. \ref{fig:ny04}.
For comparison, we use the same cell-size here.  High neutrino fraction enhances the electrons to satisfy the chemical equilibrium (see $\mu_e-\mu_\nu$ in Eq.~(\ref{eq:02})). 
As a result, the Coulomb potential changes drastically at the trapped-neutrino case as shown in this figure.

We do not present the pressure or EOS at the neutrino-trapped case as in the case of finite temperature because, at the realistic  situations in proto-neutron stars as $T=$40 MeV, $Y_l(\equiv Y_e+Y_{\nu_e})=0.4$, the mixed phase becomes unstable. 
Related  to this discussion, the recent study suggested a new possible mixed phase for the neutrino trapped matter \cite{pag}. They imposed the local charge neutrality 
a priori instead of directly solving the Poisson equation, infered from our previous result in Fig. \ref{fig:ny03}.
In such case, the surface tension and the Coulomb energy do not work to make any pasta structure, 
and the mixed phase may resemble the amorphous phase of quarks and hadrons.
Our results look to show an establishment of the local charge neutrality, but details will be discussed elsewhere.

\section{Summary and concluding remarks}
\label{sum}

We have discussed general features of the phase transitions in compact stars, 
where multi-component substance are in equilibrium. These phase transitions are non-congruent, 
and the usual Maxwell construction cannot be applied any more. 
Using the bulk calculation, 
we have presented basic concepts of the non-congruent transition. 
Although these concepts are helpful in discussing and analyzing the properties of the phase transitions 
in compact stars, we must bear in mind that they are obtained in an idealized situation and must be carefully applied 
in realistic calculations by taking into account rather complicated effects such as the finite-size effects. 
Actually we found that uniform nuclear matter can exist in the mechanically unstable region.

When the finite-size effects are taken into account, 
the structured mixed phase (pasta phase) emerges to form 
the crystalline structures. 
Charge screening effect and the rearrangement effect of particle densities in the presence of 
the Coulomb interaction is also emphasized; 
they sometimes cause an instability of the geometrical structures in the mixed phase. 
In the extreme case the EOS resembles the one given by the Maxwell construction.

Liquid-gas transition in asymmetric nuclear matter and hadron-quark deconfinement transition 
are studied and their mixed phases have been calculated by using the Thomas-Fermi approximation for particle densities and 
the Wigner-Seitz (WS) cell approximation for the pasta structure. 
We have shown many interesting aspects of these phase transitions. 
However, we have also seen that the maximum mass of hybrid stars is rather low  
in the light of a recent observation \cite{dem}, since contamination of hyperons considerably softens EOS. 
The transition to quark matter suppresses their appearance, 
but the maximum mass is still low due to the rather soft EOS of quark matter. 
So, we need another idea about EOS's of hadrons and quarks to circumvent this situation.

So far, most works on the pasta structures have used the WS cell 
with ansatz about the geometrical structures like droplet, rod, slab and so on.
If we want to see the possibility of the intermediate shapes, or study the mixed phase 
in a more realistic way, 
a study is desirable without the WS cell approximation or ansatz about the geometrical structures.
Recently we have performed a three-dimensional calculation of non-uniform nuclear matter
based on the relativistic mean-field model and the Thomas-Fermi approximation \cite{oka}.
Introducing a cubic cell with periodic boundary conditions and discretizing it into grid, 
we numerically solve the coupled field equations for meson mean-fields and the Coulomb potential.
Randomly distributing fermions (n, p, e) on the grid points as an initial condition, 
we relax their density distributions to attain the uniformity of the chemical potentials.
As a preliminary result we demonstrate that pasta structures, previously given by the WS cell approximation, 
also appear in the three-dimensional calculation.

Formation of nuclear pasta has been discussed in supernovae \cite{wata09} 
or possible observation of the signals of the mixed phase has been suggested 
in the spectra of the gravitational waves \cite{sot}.
The phenomenological implications of the pasta structures should deserve more elaborate studies;  
their thermodynamical or dynamical property is interesting; e.g. 
the neutrino opacity or the thermal conductivity may be affected by the pasta structures, 
and their elasticity may cause a discontinuous change of internal structure of compact stars.
  In particular 
the crystalline structure in the core region may bring about 
novel dynamical effects on compact-star phenomena  
by way of deformation or oscillation.

\printindex

\begin{thebibliography}{1}

\bibitem{sha} S.~L.~Shapiro and S.~A.~Teukolsky, {\it Black Holes, White Dwarfs, and Neutron Stars},
John Wiley \& Sons Inc, 1983.

\bibitem{tama} T. Kunihiro, T. Muto, T. Takatsuka, R. Tamagaki, T. Tatsumi, Prog. Theor. Phys. Suppl. {\bf 112} (1993).  

\bibitem{glen} N.K. Glendenning, {\it Compact Stars}, Springer, 2000.

\bibitem{hae} P. Haensel, A.Y. Pothekhin and D.G. Yakovlev, {\it Neutron Stars 1: Equation of State and Structure}, Springer, 2007.


\bibitem{bec}
             W. Becker ed., {\it Neutron Stars and Pulsars}, Springer, 2009.

\bibitem{sag}
             I. Sagart et al., Phys. Rev. Lett. {\bf 102} (2009) 081101.

\bibitem{wata09}
G. Watanabe, T. Maruyama, K. Sato, K. Yasuoka and T. Ebisuzaki, Phys. Rev. Lett. {\bf 94} (2005) 031101.\\
G. Watanabe, H. Sonoda, T. Maruyama, K. Sato, K. Yasuoka, and T. Ebisuzaki, Phys. Rev. Lett. {\bf 103} (2009) 121101.              


\bibitem{sot}
H. Sotani, N. Yasutake, T. Maruyama and T. Tatsumi, Phys. Rev. {\bf D} 83 (2011) 024014.

\bibitem{revpasta}
T.~Maruyama, T.~Tatsumi, T.~Endo, and S.~Chiba,
 Recent Res.\ Devel.\ in Physics {\bf 7}, 1 (2006).

\bibitem{Rav83} 
D.~G.~Ravenhall, C.~J.~Pethick and J.~R.~Wilson,
{\it Phys. Rev. Lett.} {\bf 27} (1983) 2066.

\bibitem{gle92}
N.K. Glendenning, Phys. Rev. {\bf D46} (1992) 1274.\\
N.K. Glendenning, Phys. Rep. {\bf 342} (2001) 393.

\bibitem{mag96}
W.R. Magro, D.M. Ceperly, C. Pierleoni, and B. Bernu,
Phys. Rev. Lett. {\bf 76} (1996) 1240.

\bibitem{ron}
C.R. Ronchi, I.L. Iosilevski, E.S. Yakub, {\it Equation of State of Uranium Dioxide}, Springer, 2004.

\bibitem{khra06}
S.A. Khrapak et al.,
Phys. Rev. Lett. {\bf 96} (2006) 015001.

\bibitem{for07}
V.E. Fortov et al.,
Phys. Rev. Lett. {\bf 99} (2007) 185001.

\bibitem{net08}
N. Nettelmann, B. Holst, A. Kietzmann, M. French, R. Redmar and D. Blaschke,
Astrophys. J. {\bf 683} (2008) 1217.


\bibitem{ios} 
I. Iosilevskiy, Talk at Int. Congress on Plasma Physics, Fukuoka, Japan, Sept. 2008.

\bibitem{HPS93}
H. Heiselberg, C.J. Pethick and E.F. Staubo, Phys. Rev. Lett. {\bf
70} (1993) 1355.

\bibitem{voskre}
D.~N.~Voskresensky, M.~Yasuhira and T.~Tatsumi,
{\it Phys. Lett.} {\bf B541} (2002) 93; {\it Nucl. Phys.}
{\bf A723} (2003) 291.

\bibitem{marukaon06}
T. Maruyama, T.~Tatsumi, V.~N.~Voskresensky, T.~Tanigawa, and S.~Chiba,
Phys, Rev. {\bf C73} (2006) 035802.

\bibitem{end05}
T.~Endo, T.~Maruyama, S.~Chiba, and T.~Tatsumi,
 Nucl.\ Phys.\ {\bf A749} (2005) 333;

 T.~Endo, T.~Maruyama, S.~Chiba, and T.~Tatsumi,
 Prog.\ Theor.\ Phys.\ {\bf 115} (2006) 337;

\bibitem{hyp07}
T. Maruyama, S. Chiba, H.-J. Schulze and T. Tatsumi,
Phys. Rev. {\bf D76} (2007) 123015; Phys. Lett. {\bf B659} (2008) 192.


\bibitem{yas09}
N. Yasutake,T. Maruyama, T. Tatsumi, Phys. Rev. {\bf D80} (2009) 123009; in preparation (2011).

\bibitem{tester}
J.W.Tester and M. Mondell,
{\it Thermodynamics and its applications}, Prentice Hall PTR, 1997.


\bibitem{chom04}
P. Chomaz, C. Colonna and J. Randrup,
Phys. Rep. {\bf 389} (2004) 263.

\bibitem{lan}
L.D. Landau and E.M. Lifshitz,
{\it Statistical Physics}, Elsevier, 2007.


\bibitem{avancini}
S. S. Avancini, L. Biro, Ph. Chommaz, D. P. Menezes, and C. Provid\~encia,
Phys. Rev. C {\bf 74} (2006) 24317.


\bibitem{friedman}
B. Friedman and V. R. Pandharipande,
Nucl. Phys. A {\bf 361} (1981) 502.

\bibitem{stan}
H.E. Stanley,
{\it Introduction to Phase Transitions and Critical Phenomena}, Oxford U. Press, 1971.


\bibitem{pet95}
C. Pethick, D.G. Ravenhall and C.P. Lorenz,
Nucl. Phys. {\bf A584} (1995) 675.

\bibitem{muller95}
H. M\"uller and B.D. Serot,
Phys. Rev. {\bf C52} (1995) 2072.

\bibitem{marg03}
J. Margueron and P. Chomaz,
Phys. Rev. {\bf C67} (2003) 041602(R).

\bibitem{prov06}
C. Provid\~encia, L. Brito, S.S. Avancini, D.P. Menezes and Ph. Chomaz,
Phys. Rev. {\bf C73} (2006) 025805.\\
S.S. Avancini, L. Brito, Ph. Chomaz, D.P. Menezes and C. Provid\~encia,
Phys. Rev. {\bf C74} (2006) 024317.\\
L. Brito, C. Provid\~encia, A.M. Santos, S.S. Avancini, D.P. Menezes and P. Chomas,
Phys. Rev. {\bf C74} (2006) 045801.

\bibitem{avanc08}
S.S. Avancini, D.P. Menezes, M.D. Alloy, J.M. Marinelli, M.M.W. Moraes and C. Provid\~encia,
Phys. Rev. {\bf C78} (2008) 015802.

\bibitem{rand09}
J. Randrup,
Phys. Rev. {\bf C79} (2009) 054911.

\bibitem{pag}
G. Pagliara, M. Hempel and J. Schaffner-Bielich, Phys. Rev. Lett. {\bf 103} (2009) 171102.\\
M. Hempel, G. Pagliara, J. Schaffner-Bielich, Phys. Rev. {\bf D80} (2009) 125014.


\bibitem{Baym71}
G. Baym, H. A. Bethe, and C. J. Pethick, Nucl. Phys. {\bf A175} (1971) 225.



\bibitem{Has84} 
M.~Hashimoto, H.~Seki and M.~Yamada,
{\it Prog. Theor. Phys.} {\bf 71} (1984) 320.


\bibitem{Oya93} 
K.~Oyamatsu,
{\it Nucl. Phys.} {\bf A561} (1993) 431.

\bibitem{Lor93} 
C.~P.~Lorentz, D.~G.~Ravenhall and C.~J.~Pethick,
{\it Phys. Rev. Lett.} {\bf 25} (1993) 379.

\bibitem{Cheng97}
K.~S.~Cheng, C.~C.~Yao and Z.~G.~Dai,
{\it Phys. Rev.} {\bf C55} (1997) 2092.

\bibitem{she98}
H. Shen, H. Toki, K. Oyamatsu and K. Sumiyoshi, Nucl. Phys. {\bf A637}, 435 (1998).

\bibitem{goge}
P. G\"ogelein, E.N.E. van Dalen, C. Fuchs and H. M\"uther,
Phys. Rev. {\bf C76} (2007) 024312.\\
P. G\"ogelein and H. M\"uther,
Phys. Rev. {\bf C77} (2008) 025802.

\bibitem{nak}
K. Nakazato,K. Oyamatsu, S. Yamada, Phys. Rev. Lett. {\bf 103} (2009) 132501.

\bibitem{Aic91} 
J.~Aichelin, Phys. Rep. {\bf 202} (1991) 233;
and references cited therein.

\bibitem{Mar98}
T.~Maruyama, K.~Niita, K.~Oyamatsu, T.~Maruyama, S.~Chiba and A.~Iwamoto,
{\it Phys. Rev.}  {\bf C57} (1998) 655.
T.~Kido, T.~Maruyama, K.~Niita and S.~Chiba,
{\it Nucl. Phys.} {\bf A663}-{\bf 664} (2000) 877.


\bibitem{Gen00}
G.~Watanabe, K.~Sato, K.~Yasuoka and T.~Ebisuzaki,
{\it Phys. Rev.}  {\bf C66} (2002) 012801.\\
G. Watanabe, Prog. Thor. Phys. Suppl. {\bf 186} (2010) 45.  

\bibitem{wata04}
G. Watanabe, K. Sato, K. Yasuoka and T. Ebisuzaki, Phys. Rev. {\bf C69} (2004) 055805.



\bibitem{Wil85} 
R.~D.~Williams and S.~E.~Koonin,
{\it Nucl. Phys.} {\bf A435} (1985) 844.


\bibitem{oka}
M. Okamoto et al., in preparation, 2011.

\bibitem{refDFT}
{\it Density Functional Theory}, ed. E.~K.~U.~Gross and R.~M.~Dreizler,
Plenum Press ,1995.


\bibitem{par}
R.G. Parr and W. Yang,
{\it Density-Functional Theory of Atoms and Molecules}, Oxford U. Press, 1989.


\bibitem{maru05}
T. Maruyama, T. Tatsumi, D.N. Voskresensky, T. Tanigawa and S. Chiba, 
Phys. Rev. {\bf C72} (2005) 015802.


\bibitem{ARRW}
M. Alford, K. Rajagopal, S. Reddy and F. Wilczek, Phys.Rev.
{\bf D64} (2001) 074017.





\bibitem{glenpei}
N.K. Glendenning and S. Pei, Phys. Rev. {\bf C52} (1995) 2250.

\bibitem{QCD}
               T. Schaefer, arXiv:hep-ph/0509068.\\
               P. Braun-Munzinger and J. Wambach, Rev. Mod. Phys. {\bf 81} (2009) 1031.\\
               K. Fukushima and T. Hatsuda, arXiv:1005.4814 
\bibitem{uni}
            D. Boyanovsky, H.J. de Vega, D.J. Schwarz,
	Ann. Rev. Nucl. Part. Sci. (2006) 441. 




\bibitem{nemura09}
H. Nemura, N. Ishii, S. Aoki, and T. Hatsuda, Phys. Lett. {\bf B673}, (2009) 2.

\bibitem{baldo98a}
{\it Nuclear Methods and the Nuclear Equation of
State}, M. Baldo, World Scientific, Singapore, (1999).

\bibitem{baldo98b}
M. Baldo, G. F. Burgio, and H.-J. Schulze, Phys. Rev. {\bf C58}, (1998) 3688.

\bibitem{schulze95}
H.-J. Schulze et al., Phys. Lett. {\bf B 355}, (1995) 21.

\bibitem{vidana95}
I. Vida\~na et al., Phys. Rev. {\bf C 61}, (2000) 025802.

\bibitem{wiringa95}
R. B. Wiringa, V. G. J. Stoks, and R. Schiavilla, Phys. Rev. {\bf C 51}, (1995) 38.

\bibitem{pudliner97}
B. S. Pudliner et al., Phys. Rev. {\bf C 56}, (1997) 1720.

\bibitem{maessen89}
P. M. M. Maessen, T. A. Rijken, and J. J. de Swart, Phys.Rev. {\bf C 40}, (1989) 2226.

\bibitem{baldo99}
M. Baldo and L. S. Ferreira, Phys. Rev. {\bf C 59}, (1999) 682. 

\bibitem{nicotra06a}
O. E. Nicotra, M. Baldo, G. F. Burgio, and H.-J. Schulze, Astron. Astrophys. {\bf 451}, (2006) 213. 

\bibitem{nicotra06b}
O. E. Nicotra, M. Baldo, G. F. Burgio, and H.-J. Schulze, Phys. Rev. {\bf D 74}, (2006) 123001.

\bibitem{baldo04}
M. Baldo, L. S. Ferreira, and O. E. Nicotra, Phys. Rev. {\bf C 69}, (2004) 034321.

\bibitem{jaf}
E. Farhi and R. L. Jaffe, Phys. Rev. {\bf D 30}, (1984) 2379. 

\bibitem{latt}
K. Kajantie,	L. Ka\"{r}ka\"{i}nen, and K. Rummukainen, Nucl. Phys. {\bf B357}, (1991) 693.

\bibitem{nsmass}
For a recent review, see
 D. Page and S. Reddy,
 Annu. Rev. Nucl. Part. Sci. {\bf 56}, 327 (2006).


\bibitem{dem}
P.B. Demorest et al., Nature {\bf 467} (2010) 1081.

\end{thebibliography}
\end{document}